\documentclass{aa}
\usepackage{graphicx}
\usepackage{multirow}
\usepackage{txfonts}
\usepackage{xspace}
\usepackage{xcolor}
\usepackage{url}
\usepackage[colorlinks=true, allcolors=blue]{hyperref}
\usepackage[flushleft]{threeparttable}
\usepackage{orcidlink}
\newcommand{\src}{2S~1417$-$624\xspace}

\newcommand{\ixpe}{\text{IXPE}\xspace}

\begin{document}

\title{Probing the emission geometry of the X-ray pulsar 2S~1417$-$624 during a weak outburst with NICER, IXPE, and NuSTAR}

\author{M.~Zhou \inst{\ref{in:Tub}}\orcidlink{0000-0001-8250-3338}
\and P.-J.~Wang \inst{\ref{affil:QHU1}, \ref{affil:QHU2}}\orcidlink{0000-0002-6454-9540}
\and H.-H.~Liu \inst{\ref{in:Tub}}\orcidlink{0000-0003-2845-1009}
\and L.~Ducci \inst{\ref{in:Tub}}\orcidlink{0000-0002-9989-538X}
\and S.~S.~Tsygankov \inst{\ref{in:UTU}, \ref{in:Tub},\ref{in:IHEP}}\orcidlink{0000-0002-9679-0793}
\and Q.-C.~Zhao \inst{\ref{in:IHEP}, \ref{affil:UCAS}}\orcidlink{0000-0001-9893-8248}
\and J.~Poutanen\inst{\ref{in:UTU}}\orcidlink{0000-0002-0983-0049}
\and L.~Ji \inst{\ref{affil:SYSU}}\orcidlink{0000-0001-9599-7285}
\and V.~F.~Suleimanov \inst{\ref{in:Tub}}\orcidlink{0000-0003-3733-7267}
\and A.~A.~Mushtukov \inst{\ref{in:Oxford}}\orcidlink{0000-0003-2306-419X}
\and Q.~Liu \inst{\ref{affil:HEBTU}}\orcidlink{0000-0002-4132-5720}
\and C.~M.~Diez \inst{\ref{affil:ESAC}}\orcidlink{0000-0001-6520-4600}
\and L.~Abalo \inst{\ref{affil:Leiden}}\orcidlink{0009-0004-0573-9637}
\and V.~Grinberg \inst{\ref{in:ESTEC}}\orcidlink{0000-0003-2538-0188}
\and A.~Santangelo \inst{\ref{in:Tub}}\orcidlink{0000-0003-4187-9560}
}

\institute{
Institut f\"ur Astronomie und Astrophysik, Universit\"at T\"ubingen, Sand 1, D-72076 T\"ubingen, Germany \label{in:Tub} \\ \email{wangpj@qhu.edu.cn, honghui.liu@uni-tuebingen.de}
\and Research Center of Astronomy, QingHai University, Xining, 810016, China \label{affil:QHU1}
\and Department of Physics and Astronomy, QingHai University, Xining, 810016, China \label{affil:QHU2}
\and Department of Physics and Astronomy, 20014 University of Turku, Finland \label{in:UTU}
\and Key Laboratory of Particle Astrophysics, Institute of High Energy Physics, Chinese Academy of Sciences, Beijing 100049, China\label{in:IHEP} 
\and University of Chinese Academy of Sciences, Chinese Academy of Sciences, Beijing 100049, China \label{affil:UCAS}
\and School of Physics and Astronomy, Sun Yat-Sen University, Zhuhai 519082, China \label{affil:SYSU}
\and Astrophysics, Department of Physics, University of Oxford, Denys Wilkinson Building, Keble Road, Oxford OX1 3RH, UK \label{in:Oxford}
\and School of Physics, Hebei Normal University, Shijiazhuang 050024, China \label{affil:HEBTU}
\and European Space Agency (ESA), European Space Astronomy Centre (ESAC), Camino Bajo del Castillo s/n, 28692 Villanueva de la Cañada, Madrid, Spain \label{affil:ESAC}
\and Huygens-Kamerlingh Onnes Laboratory, Leiden University, Postbus 9504, 2300 RA Leiden, The Netherlands \label{affil:Leiden}
\and European Space Agency (ESA), European Space Research and Technology Centre (ESTEC), Keplerlaan 1, 2201 AZ Noordwijk, The Netherlands \label{in:ESTEC}
}
  
\titlerunning{Probing the emission geometry of 2S~1417$-$624 during a weak outburst with NICER, IXPE, and NuSTAR}
\authorrunning{M.~Zhou et al.}

\date{-- / --}

\abstract{We report results from a multi-mission observational campaign of the transient X-ray pulsar 2S~1417$-$624 during its 2025 outburst, using data from NICER, IXPE, and NuSTAR. Phase-averaged and phase-resolved spectroscopy with NICER and NuSTAR reveal that a typical broken power-law model with a high-energy cut-off well describes the broadband spectra. Several spectral parameters, however, show clear and systematic modulations with pulse phase, indicating variations in the physical conditions of the emitting plasma over the neutron star's rotation. IXPE provides the first polarimetric measurements of this source, yielding a phase-averaged polarization degree (PD) of $4.8 \pm 1.2$\% and a polarization angle (PA) of ${17}\degr \pm {7}\degr$, both quoted at the $1\sigma$ confidence level. Fitting the phase-resolved PA with the rotating vector model (RVM) gives a magnetic obliquity of $\theta = 69_{-29}^{+13}$\,deg, indicating a significantly inclined magnetic geometry that may approach a quasi-orthogonal configuration. In addition, using the unbinned photon-by-photon method, we obtain a PD of $5.9 \pm 1.2$\% across the pulse phase, together with a pulsar geometry consistent with that inferred from the binned analysis, assuming the variable PA predicted by the RVM. A simultaneous RVM fit across the three energy bands, 2--5\,keV, 5--6\,keV, and 6--8\,keV, provides the strongest constraints on the geometrical parameters, yielding $\theta = {84}_{-6}^{+4}$\,deg. Together, these findings demonstrate pronounced phase-dependent spectral and polarization variability, offering valuable constraints on the geometry and emission processes within the accretion region of this transient X-ray pulsar. }

\keywords{accretion, accretion disks -- magnetic fields -- polarization -- pulsars: individual: 2S~1417$-$624 -- stars: neutron -- X-rays: binaries}

\maketitle
\nolinenumbers 

\section{Introduction}\label{sect:intro}

Accreting X-ray pulsars (XRPs) are binary systems consisting of a strongly magnetized neutron star ($B\sim$${10}^{12}$\,G) and a companion star. The neutron star accretes material from its companion, releasing gravitational energy and forming highly concentrated emission regions near its magnetic poles \citep[for a review, see, e.g.,][]{Mushtukov_Tsygankov_2024_Review}. Constrained by the magnetic field, the infalling matter is channeled along field lines toward the poles, where it forms a hot plasma in the polar cap. The neutron star's rotation modulates this emission, producing periodic X-ray pulses. The beam pattern depends on the geometry of the emitting regions and the luminosity: in the sub-critical regime, radiation is predominantly emitted along the magnetic axis (``pencil beam''), whereas in the super-critical regime, radiation escapes laterally below the radiation-dominated shock, forming a ``fan beam'' \citep{Basko_1975, Becker_2012, Mushtukov_2015}. Transitions between these patterns manifest in changes in the pulse profile, pulsed fraction, and spectral properties, highlighting the importance of broadband, long-term monitoring for constraining accretion geometry and emission physics \citep{Reig_2013, Kong_2021, Wang_2022}. 

Recent advances in X-ray polarimetry have opened a new avenue for probing accreting XRPs. Measurements of the energy and phase dependence of the polarization degree (PD) and polarization angle (PA) can tightly constrain the emission geometry. While PD values in the X-ray band are generally modest (a few to $\sim$10\%), PA often exhibits a systematic variation with pulse phase. Modeling such variations with the rotating vector model (RVM) can yield the relative orientations of the magnetic axis, spin axis, and the observer's line of sight \citep{Poutanen_2020_RVM, Poutanen_2024_Galaxies}. Combined with spectral and pulse profile analyses, polarization measurements can provide direct observations of radiation from the accretion column and probe the neutron star's magnetic field structure \citep{Krawczynski_2019, Doroshenko_2022_HerX1}. 

\src is a prototypical transient Be/X-ray binary, discovered by SAS-3 in 1978 \citep{Apparao_1980}. It has an orbital period of $\sim$42.12\,d, and an eccentricity $e$ of $\sim$0.446  and exhibits coherent X-ray pulsations with a period of $\sim$17.64\,s  \citep{Kelley_1981, Finger_1996}. The optical companion is a B1~Ve star \citep{Grindlay_1984}. \textit{Gaia} measurements place the system at a distance of ${9.9}_{-2.4}^{+3.1}$\,kpc using the Data~Release~2 \citep{Bailer-Jones_2018} and ${7.4}_{-1.8}^{+3.1}$\,kpc using the Data~Release~3 \citep{Bailer-Jones_2021}, whereas torque modeling suggests a significantly larger distance of $\sim$20\,kpc \citep{Ji_2020}. The surface magnetic field strength of the neutron star is estimated to be $\sim$$9 \times 10^{12}$\,G, inferred from the cyclotron resonant scattering feature (CRSF) detected at $\sim$100\,keV \citep{LiuQ_2024}. The source has undergone several bright type-II outbursts in 1994, 1999, 2009, 2018, and 2021, during which the pulse profile, pulsed fraction, and their luminosity dependence show pronounced evolution. For example, the profile can evolve from double- to triple- or quadruple-peaked, and in some cases, the pulsed fraction anti-correlates with luminosity \citep{Inam_2004, Gupta_2018, Gupta_2019, Ji_2020, LiuQ_2024, LiuQ_2025}. In addition to these bright events, the source exhibits lower-luminosity type-I outbursts near periastron, which are shorter in duration and significantly fainter \citep{Finger_1996}. Systematic studies of type-I outbursts are scarce, with only a few works (e.g., \citealt{Inam_2004}) analyzing their pulse and spectral properties using 1999--2000 RXTE data. 

In this work, we present an analysis of the timing and spectral characteristics of \src during a new outburst observed in 2025, and report the first X-ray polarization measurement for this source. The remainder of this paper is organized as follows. Section~\ref{sect:data} describes the long-term behavior of the source and provides an overview of the data sets analyzed. The main results are presented in Sect.~\ref{sect:results}, followed by a discussion in Sect.~\ref{sect:discussion}. Finally, Sect.~\ref{sect:summary} summarizes our findings and offers an outlook for future work. 

\section{Observations and data reduction}\label{sect:data}

\begin{figure}
\centering
\includegraphics[width=0.99\linewidth]{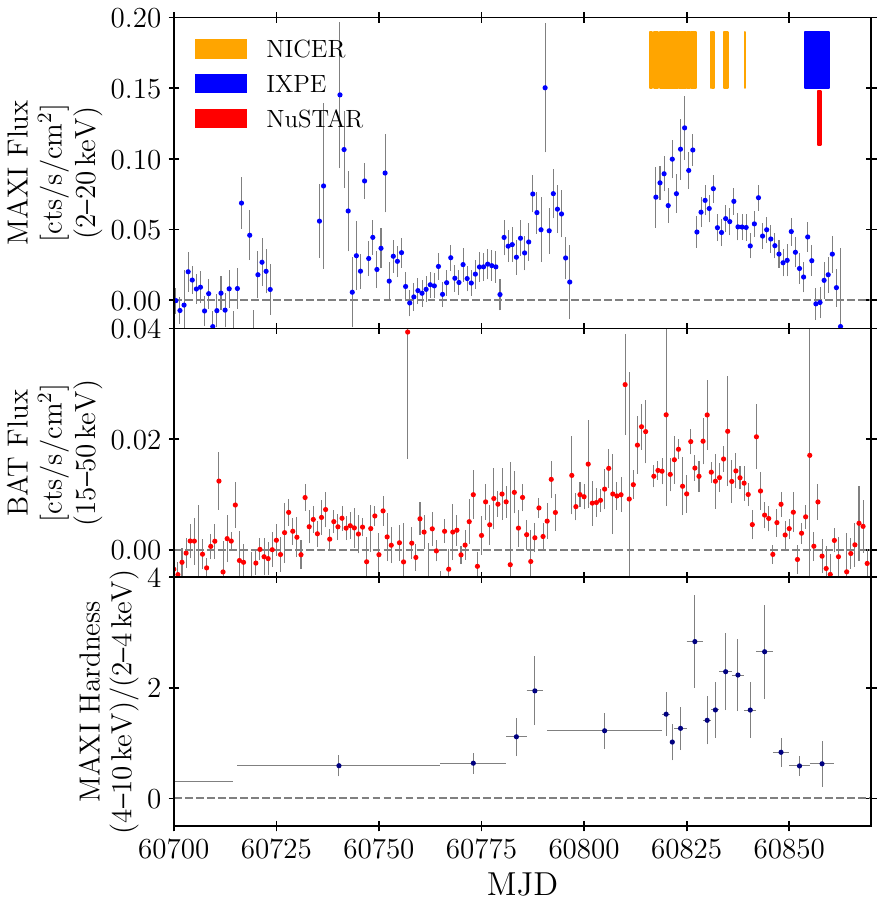}
\caption{Light curves of \src{} during its 2025 outburst observed with MAXI and Swift/BAT. The lower panel shows the source hardness, defined as the ratio of the MAXI photon flux in the 4--10\,keV band to that in the 2--4\,keV band, rebinned to achieve a significance of at least 3$\sigma$. The NICER, IXPE, and NuSTAR observation periods are indicated by orange, blue, and red bars, respectively. }
\label{fig:longterm-lc}
\end{figure}

\src{} underwent a new type-II outburst in April 2025 that was significantly weaker than previous giant outbursts from the source, as revealed by continuous monitoring with all-sky X-ray instruments. Figure~\ref{fig:longterm-lc} shows the long-term light curves from MAXI\footnote{\url{http://maxi.riken.jp/}}~\citep{Matsuoka_2009_MAXI} in the 2--20\,keV band and Swift/BAT\footnote{\url{https://swift.gsfc.nasa.gov/results/transients/}}~\citep{Barthelmy_2005_BAT} in the 15--50\,keV band. The maximum brightness reached $\sim$70\,mCrab in the BAT band around MJD\,60825, indicating a significant enhancement in the hard X-ray flux compared to the quiescent level. Around the outburst peak, the source also became harder, as indicated by an increase in the MAXI hardness ratio between the 4--10 and 2--4\,keV bands. After the peak, the source entered a slow decay and eventually returned to its pre-outburst flux level. The entire event lasted for roughly two months. 

The outburst was observed with the Neutron Star Interior Composition Explorer (NICER; \citealt{Gendreau_2016}), the Imaging X-ray Polarimetry Explorer (\ixpe; \citealt{Weisskopf_2022}), and the Nuclear Spectroscopic Telescope Array (NuSTAR; \citealt{Harrison_2013_NuSTAR}). An overview of the observations from each X-ray mission, along with the corresponding data reduction procedures, is provided below. 

\subsection{NICER}\label{sect:data:nicer}

\src{} was observed with NICER between May 21 and June 13, 2025 (MJD\,60816--60839), near the peak of the outburst, with a total effective exposure time of 27.5\,ks. Data reduction was performed with \texttt{HEASoft} v6.36 and the \texttt{xti20240206} calibration database. The cleaned event lists were generated by the standard \texttt{nicerl2} task, which includes calibration and filtering steps. To minimize the background effects, we applied a custom set of selection criteria to identify good time intervals (GTIs) based on standard NICER screening criteria: an overshoot rate lower than 10\,cts per FPM (\texttt{FPM\_OVERONLY\_COUNT<10}), an undershoot rate lower than 200\,cts per FPM (\texttt{FPM\_UNDERONLY\_COUNT<200}), and a geomagnetic cutoff-rigidity greater than 1.5\,GV (\texttt{COR\_SAX>1.5}). Spectra were extracted from the resulting GTIs using the tool \texttt{nicerl3-spect}. The background spectrum was estimated with model \texttt{SCORPEON}. The energy band for the spectral analyses is 0.7--10\,keV. 

\subsection{IXPE}\label{sect:data:ixpe}
\ixpe is a collaborative project between NASA and the Italian Space Agency, launched aboard a Falcon 9 rocket on December 9, 2021, intending to conduct imaging polarimetry in the 2--8\,keV energy range \citep{Weisskopf_2022}. The observatory is composed of three identical grazing-incidence X-ray telescopes, each consisting of a mirror module and a polarization-sensitive detector unit (DU) that employs a gas pixel detector \citep{Soffitta_2021, Baldini_2021}. For every detected photon, \ixpe records its sky position, arrival time, and energy, as well as the emission direction of the associated photoelectron, which is essential for deriving polarization information. 

\ixpe observed \src{} from June 27 to July 3, 2025 (MJD\,60853--60859), during which the source was in the final phase of its outburst, with a total effective exposure of 320\,ks. We used data from all three DUs, including the reprocessed DU2 data provided by the IXPE team, which corrects for the pixel failures that occurred in April 2025, around MJD\,60779. 

The data were processed using the \texttt{HEASoft} package together with calibration files from the calibration database released on April 27, 2026. Before analysis, position offset corrections and energy calibration were applied. Source photons were extracted from a circular region of radius $R_\text{src}=70\arcsec$ centered on the target, while the background was estimated from an annular region around the source, defined by inner and outer radii of $R_\text{in}=120\arcsec$ and $R_\text{out}=240\arcsec$, respectively. In the 2--8\,keV energy range, the total source count rate from all the three DUs is approximately $0.3\,\mathrm{cps}$, whereas the average background rate in the same band is about $0.001\,\mathrm{cps}\ \mathrm{arcmin}^{-2}$ per DU. Given the relatively low count rate, we perform both background rejection using version 3.2 of the rejection script and background subtraction in the polarimetric analysis, following the recommendations of \citet{DiMarco_2023} and \citet{DiMarco_2026}. 

The Stokes $I$, $Q$, and $U$ spectra for the three DUs were produced using the \texttt{xselect} tool, adopting the weighted method \texttt{NEFF} to improve the statistical significance \citep{DiMarco_2022}. The corresponding ancillary response files are generated with \texttt{ixpecalcarf}, using the proper data files in the calibration database. The $I$ spectra were then re-binned to ensure a minimum of 30 counts per energy channel, and the $Q$, $U$ spectra are re-binned using a constant energy bin width of 120 eV, as suggested by \citet{DiMarco_2026}. The resulting spectra were fitted simultaneously with the \textsc{xspec} package (version 12.15.1; \citealt{Arnaud_1996}). Fits were performed using $\chi^2$ statistics, and uncertainties are quoted at the 68.3\% confidence level (1$\sigma$), unless stated otherwise. 

We also applied the tool \texttt{ixpepolarization} from the \texttt{HEASoft} package to carry out a model-independent polarimetric analysis. This approach bypasses the use of the full energy response matrix but incorporates energy dependence by assigning weights to individual photon contributions. All photons within the chosen time interval and energy range were included, and the Stokes parameters were derived via ``unweighted'' summation, as suggested by the IXPE team. We also perform background subtraction by conducting model-independent polarimetric analysis on both the source and background regions, obtaining the unnormalized Stokes parameters, and applying a scaled subtraction to estimate the Stokes parameters and their associated uncertainties. 

\subsection{NuSTAR}
NuSTAR observed \src{} on July 1, 2025 (MJD\,60857) with observation ID~91101319002 and an exposure time of approximately 32.5~ks, overlapping with the IXPE observation. We processed the data using \texttt{nupipeline} to produce the cleaned event files, and employed \texttt{nuproducts} to extract the source and background spectra and light curves. The source region was defined as a circular aperture with a radius of $120^{\prime\prime}$ centered on the source position, while the background region was chosen as a circular aperture of the same radius located as far from the source as possible. 

\section{Results} \label{sect:results}
\subsection{Pulsar timing studies and the pulse profiles}
Before determining the precise pulse period, we first removed the effects of both the Earth's motion and the binary's orbital motion. Event arrival times were corrected to the Solar System barycenter using the \texttt{barycorr} tool from the \textsc{ftools} package. To compensate for orbital motion, we applied additional corrections using the orbital ephemerides provided by the Fermi Gamma-ray Burst Monitor\footnote{\url{https://gammaray.nsstc.nasa.gov/gbm/science/pulsars/lightcurves/2s1417.html}\label{footnote:2S1417}}\citep{Malacaria_2020_GBM}, which are based on the solutions of \citet{Finger_1996} and \citet{Inam_2004}, but include an adjustment of the orbital period by -1.25/84 days, extrapolated to the epoch of the \ixpe observation (see Table~\ref{table:orb-pars}). 

\begin{table}
\centering
\caption{Orbital parameters for 2S~1417$-$624 (ephemeris adopted from Fermi Gamma-ray Burst $\text{Monitor}^{\ref{footnote:2S1417}}$). } 
\begin{tabular}{lcc}
\hline
\hline
Parameter & Value & Unit\\
\hline
    Orbital period  & $42.175$ & d  \\
    $T_{\text{periastron}}$  & $51612.17000$ & MJD  \\
    $a_{\rm X}\sin i$  & $188$ & light-sec  \\
    Longitude of periastron  & $300.30$ & deg   \\
    Eccentricity  & $0.4460$ &    \\
\hline
\end{tabular}
\label{table:orb-pars}
\end{table}

\begin{figure}
\centering
\includegraphics[width=0.95\linewidth]{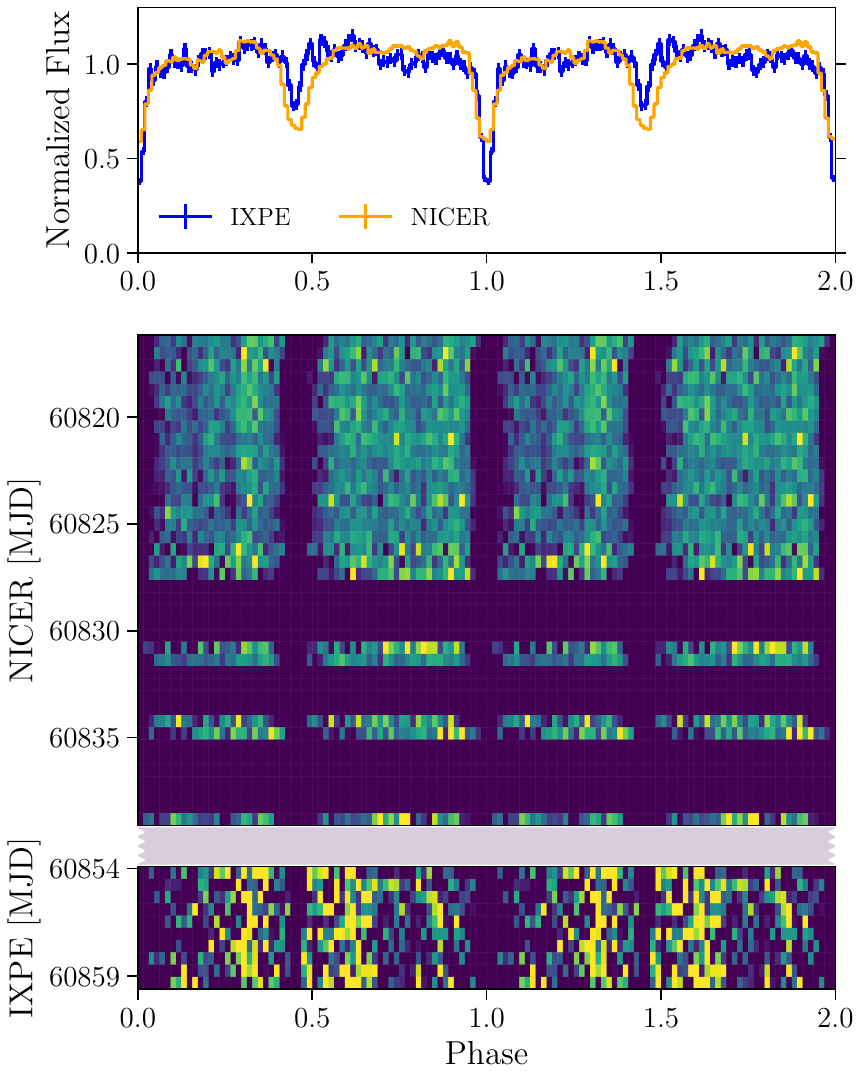}
\caption{Pulse profiles of \src{} observed by NICER and \ixpe, and their phaseograms as a function of time (MJD). Fluxes are normalized to a unit mean. Profiles use 100 phase bins, and phaseograms use 64 phase bins; the phaseograms are divided into 40 time bins for NICER and 10 for IXPE.}
\label{fig:phaseogram}
\end{figure}

The pulse period ($P$) and its first derivative ($\dot{P}$) for the NICER and \ixpe data were independently determined using $Z^2$ statistics with two harmonics \citep{Buccheri_1983}, by maximizing the $Z^2$ values from all photon events in the 0.7--10\,keV band for NICER and the 2--8\,keV band for \ixpe. The resulting spin parameters and pulse epochs are listed in Table~\ref{table:time-pars}. The zero phase of the pulse profile is defined at the phase corresponding to the minimum count rate. Figure~\ref{fig:phaseogram} shows the NICER and \ixpe pulse profiles in their respective energy bands, together with phaseograms tracing pulse profile evolution over 23 days for NICER and 6 days for \ixpe. This method naturally maintains phase alignment over extended time spans, as illustrated in Fig.~\ref{fig:phaseogram}. For the NuSTAR data, as the observation lasted less than one day, only the pulse period was determined by maximizing the $Z^2$ values in the 3--79\,keV band, as listed in the last column of Table~\ref{table:time-pars}. 

\begin{figure}
\centering
\includegraphics[width=0.85\linewidth]{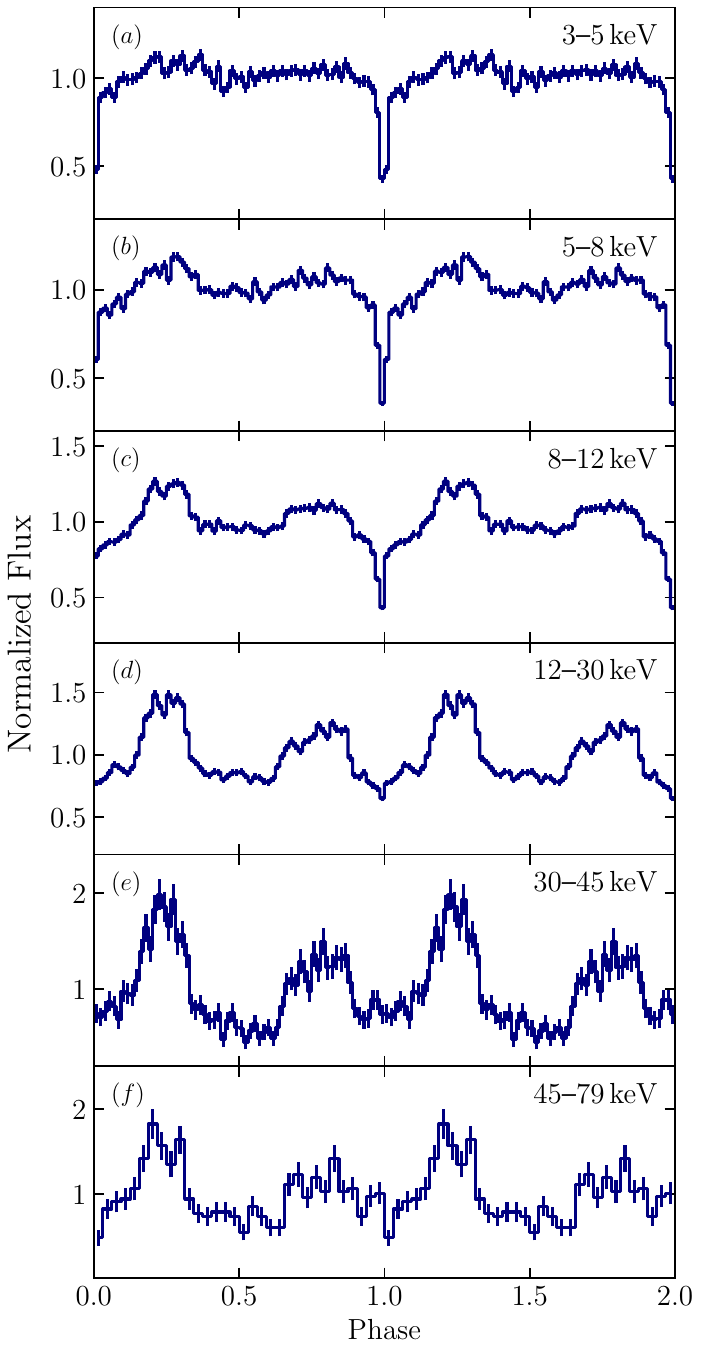}
\caption{Pulse profiles at different energies observed by NuSTAR. The fluxes are normalized such that their mean values are unity.}
\label{fig:profile-E-nustar}
\end{figure}

The pulse profiles from NICER and \ixpe are similar, each showing a prominent dip at phase zero and a shallower dip near phase $\sim$0.45. Both also display two broad peaks, each containing subtle sub-peaks. No significant changes in pulse shape are observed during either observational period. At higher energies, the pulse profile exhibits a sharper double-peaked structure than at lower energies, as indicated by the pulse profile analysis over the broad NuSTAR energy range shown in Fig.~\ref{fig:profile-E-nustar}.

\begin{table*} 
\centering
\caption{Pulsar ephemerides of \src{} with NICER, IXPE and NuSTAR. }
\begin{tabular}{lccc}
\hline
\hline
 Parameter & NICER & IXPE & NuSTAR \\
\hline
    Start [MJD] & 60816.2 & 60853.9 & 60857.0\\
    End [MJD] & 60839.1 & 60859.6 & 60857.6\\
    Exposure [ks] & 27.5 & 320.2 & 32.5 \\
    $t_0$ [MJD]\tablefootmark{$\text{a}$} & 60831.000178 & 60857.000098 & 60857.000098\\
    Pulse period  [$\mathrm{s}$] & $17.31294(1)$ & $17.307216(6)$ & $17.3072(2)$ \\
    Pulse period derivative [$10^{-9}\mathrm{s \, s^{-1}}$] & $-4.36(1)$ & $-1.38(8)$ & $|\dot{P}| \leq 7.5$ \\
\hline  
\end{tabular}
\tablefoot{
The uncertainties are reported at a 1$\sigma$ confidence level. \tablefoottext{a}{The reported reference time $t_0$ defines the zero phases of the pulse profiles (see Fig.~\ref{fig:phaseogram}). }}
\label{table:time-pars}
\end{table*}

\subsection{Spectral analysis}
\subsubsection{Phase-averaged spectral analysis}
NICER's large effective area in the soft X-ray band ($\sim$1900\,$\mathrm{cm}^{2}$ at $\sim$1.5\,keV) makes it particularly effective for probing soft X-ray emission, whereas NuSTAR's broader energy coverage, extending up to $\sim$79\,keV, provides strong constraints on the continuum slope at high energies.

\begin{figure}
\centering
\includegraphics[width=0.95\linewidth]{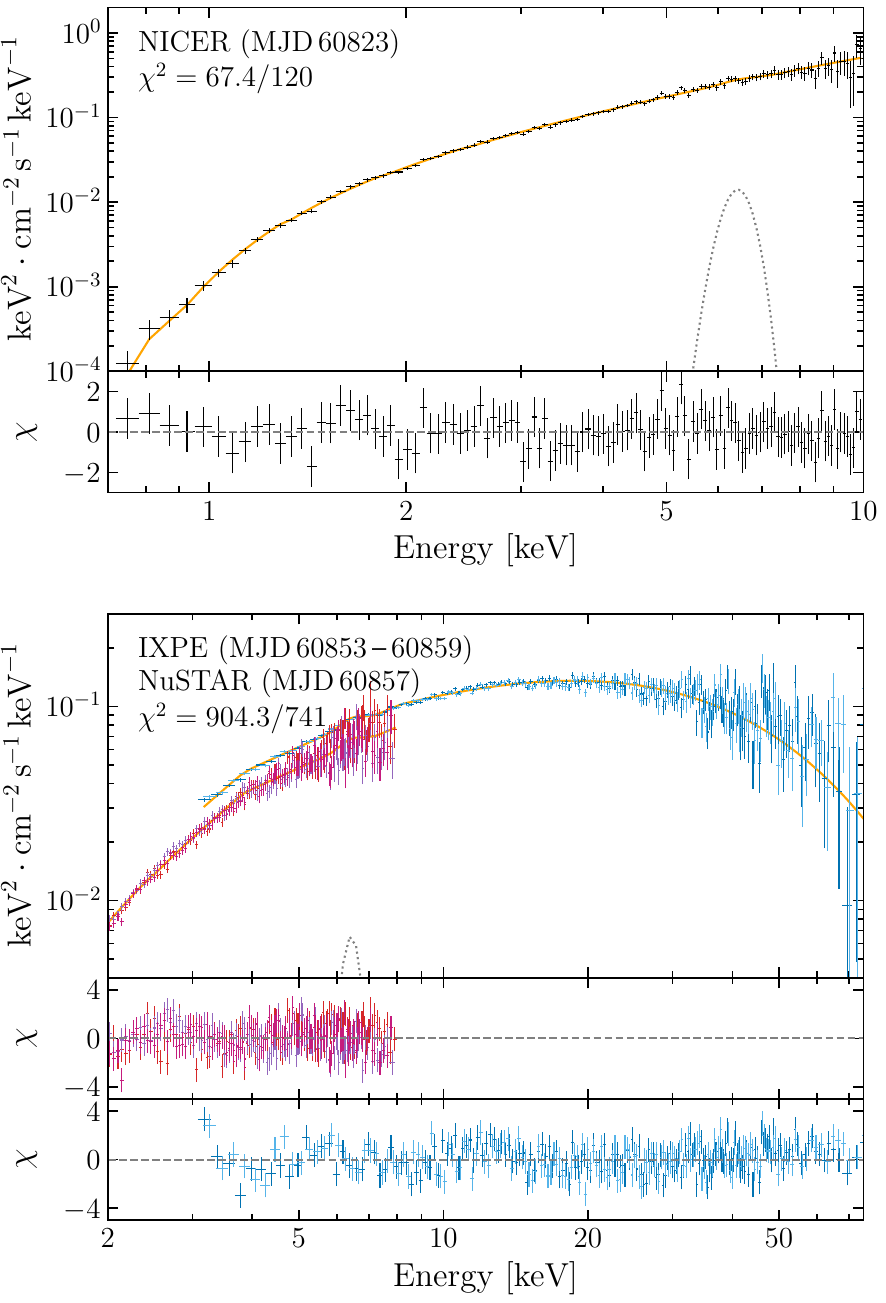}
\caption{Spectral composition of a representative phase-averaged NICER spectrum and the IXPE-NuSTAR spectra. Upper panel: The NICER data are fitted with the model $\texttt{TBabs} \times (\texttt{cutoffpl} + \texttt{gaussian})$. The solid orange line represents the best-fitting model, while the dotted grey line indicates the Gaussian component at 6.4\,keV. Lower panel: The IXPE and NuSTAR data are jointly fitted with the model $\texttt{TBabs} \times (\texttt{bknpower} \times \texttt{highecut} + \texttt{gaussian})$. The blue and cyan crosses correspond to the FPMA and FPMB data, respectively, while the red, purple, and magenta crosses correspond to the DU1, DU2, and DU3 data. The solid orange line shows the best-fitting model, and the dotted grey line represents the Gaussian component at 6.4\,keV. }
\label{fig:spectra}
\end{figure}

We adopted a systematic error of $\sim$1.5\% for all NICER/XTI spectra, as set by default in the \texttt{nicerl3-spect} tool. All NICER spectra were grouped using the optimal binning method described by \citet{Kaastra_2016}, ensuring a minimum of 30 counts per bin. 

The spectra of \src{} in the soft energy band (0.7--10\,keV) can be well described by an absorbed broken power-law, plus a Gaussian emission line at $\sim$6.4\,keV corresponding to the iron fluorescent line. However, the break energy of the broken power-law ($E_\text{break}$, typically around 4–6\,keV) is poorly constrained in the NICER spectral fits, which also prevents the high-energy photon index $\Gamma_2$ from being well determined. To obtain smoother spectra and reduce the number of free parameters, we therefore replace the broken power-law with a cutoff power-law model. In \textsc{xspec}, the model can be written as
\begin{equation*}
\texttt{TBabs} \times (\texttt{cutoffpl} + \texttt{gaussian}) \, ,
\end{equation*}
where \texttt{TBabs} accounts for photoelectric absorption by the interstellar medium, with cross-sections from \citet{Verner_1996} and elemental abundances from \citet{Wilms_2000}. The centroid energy of the \texttt{gaussian} component is fixed at 6.4\,keV, while the width parameter $\sigma$ is constrained to an upper limit of 0.3\,keV, corresponding to a narrow iron line. 

\begin{figure}
\centering
\includegraphics[width=0.95\linewidth]{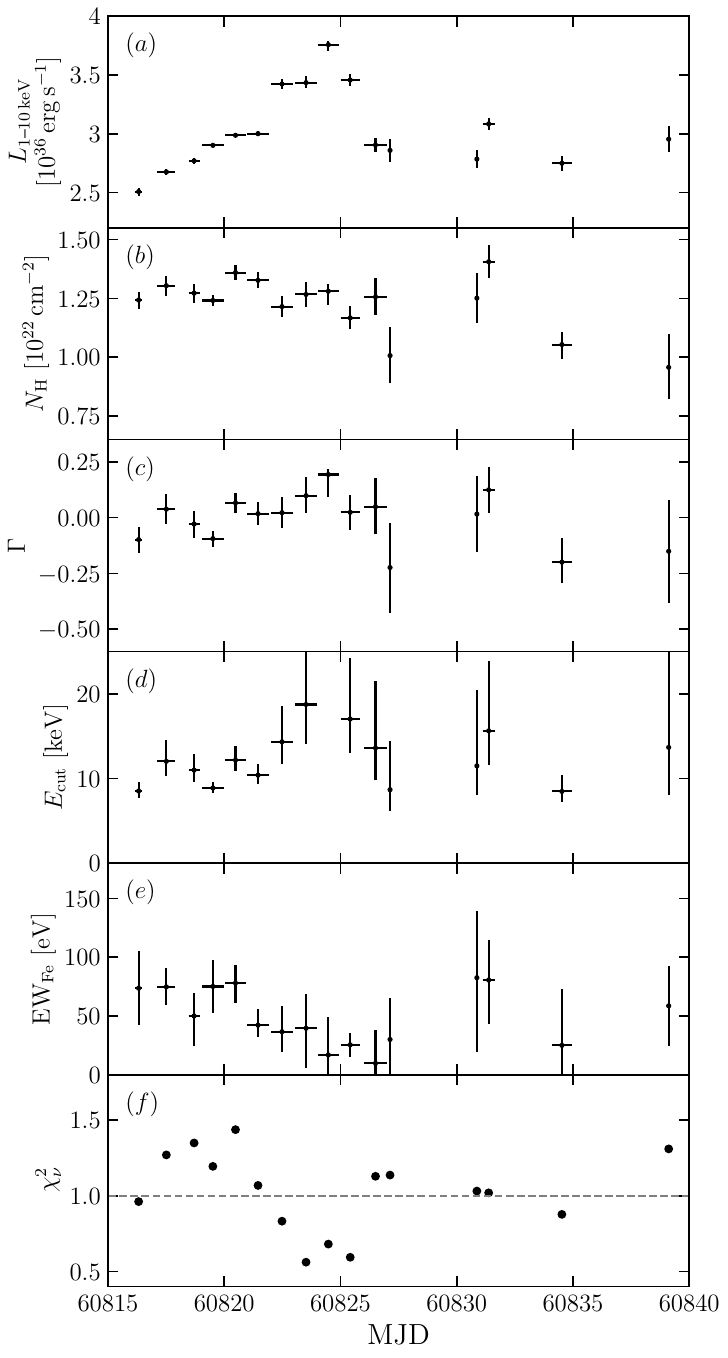}
\caption{Evolution of the key parameters of \src{} obtained from NICER phase-averaged spectral analysis as a function of time. The panels present, from ($a$) to ($f$): the 1--10\,keV luminosity assuming a distance of 7.4\,kpc; the equivalent hydrogen column density $N_\text{H}$ of the interstellar medium; the photon index $\Gamma$ and cut-off energy $E_\mathrm{cut}$ of the continuum; the equivalent width of the iron K$\alpha$ line at 6.4\,keV; and the reduced chi-square $\chi^2_\nu$ of the best-fits. }
\label{fig:phase-averaged-nicer}
\end{figure}

We first carried out spectral analysis of the phase-averaged data, which provides a more reliable representation of the spectral shape over an extended timescale. A representative spectrum is shown in the upper panel of Fig.~\ref{fig:spectra}, and the best-fit parameters resulting from the spectral analysis are presented in Fig.~\ref{fig:phase-averaged-nicer} as a function of time. No significant variations are found in the key spectral parameters during the NICER observations, except for the source luminosity, which initially increases and subsequently stabilizes. The equivalent hydrogen column density ($N_\mathrm{H}$) toward \src{} is consistently measured to be $\sim1.2\times {10}^{22}\,\mathrm{cm}^{-2}$. 

The NuSTAR spectra were grouped following the same procedure as adopted for the NICER data, and the spectral fitting was performed over the 3--79\,keV energy range. Since the absorbing column density is well constrained by the NICER observations, we fix this value at $1.2\times {10}^{22}\,\mathrm{cm}^{-2}$ for the IXPE-NuSTAR simultaneous observation. A multiplicative constant is included to account for the slight normalization differences between FPMA and FPMB. 

Owing to the broadband coverage of NuSTAR, the high-energy photon index $\Gamma_2$ can be well constrained, together with the break energy ($E_\text{break}$) of the power-law component at lower energies. To describe the full continuum, we adopt the combination $\texttt{bknpower} \times \texttt{highecut}$. A narrow iron line is also included in our model. The Stokes $I$ spectra from the simultaneous IXPE observation are incorporated in the fit to perform a joint spectral analysis. We therefore adopt the model
\begin{equation}
\texttt{constant} \times \texttt{TBabs} \times (\texttt{bknpower} \times \texttt{highecut} + \texttt{gaussian}) \, , \notag
\end{equation}
which yields a fit with $\chi^{2}/\mathrm{d.o.f.} = 904.3/741$, as shown in the lower panel of Fig.~\ref{fig:spectra}. This model is commonly used in spectral analyses of accreting XRPs \citep[see,  e.g.,][]{Hemphill_2019_4U1538, Serim_2022, Mandal_2022}. The best-fit parameters are summarized in Table~\ref{table:phase-avg-nustar}. 

\begin{table}
\centering
\caption{Best-fit spectral parameters from the phase-averaged NuSTAR and IXPE spectra with the model $\texttt{constant} \times \texttt{TBabs} \times (\texttt{bknpower} \times \texttt{highecut} \times \texttt{polconst}_{1} + \texttt{gaussian} \times \texttt{polconst}_{2})$. }
\begin{tabular}{llc}
\hline \hline
Component & Parameter & Value \\
\hline
\texttt{TBabs} & $N_{\mathrm{H}}$ $[10^{22}\mathrm{\;cm^{-2}}]$ & $1.2$ (fixed)\\
\texttt{bknpower} & $\Gamma_1$ &  $0.09 \pm 0.03 $ \\
 & $E_\text{break}$ [keV] & $3.86_{-0.07}^{+0.06}$ \\
 & $\Gamma_2$ & $1.00 \pm 0.02$ \\
\texttt{highecut} & ${E}_\mathrm{cut}$ [keV] & $8.2 \pm 0.2$\\
 & ${E}_\mathrm{fold}$ [keV] & $18.6 \pm 0.3$ \\
$\texttt{polconst}_{1}$ & PD [\%] & $4.8 \pm 1.2$ \\
 & PA [deg] & $17 \pm 7$ \\
\texttt{gaussian} & $E_\text{line}$ [keV] & 6.4 (fixed) \\
& $\textrm{EW}_\text{Fe}$ [eV] & $62 \pm 12$ \\
& $\sigma_\text{Fe}$ [keV] & $ 0.30_{-0.01}^{p} $ \\
$\texttt{polconst}_{2}$ & PD [\%] & 0 (fixed) \\
\texttt{constant} & $\mathrm{const_\text{NuSTAR/FPMA}}$ & 1 (fixed) \\
 & $\mathrm{const_\text{NuSTAR/FPMB}}$ & $0.994 \pm
  0.004$ \\
 & $\mathrm{const_\text{IXPE/DU1}}$ & $0.779 \pm 0.006$ \\
 & $\mathrm{const_\text{IXPE/DU2}}$ & $0.786 \pm 0.006$ \\
 & $\mathrm{const_\text{IXPE/DU3}}$ & $0.769 \pm 0.006$ \\
\hline
 & $\mathrm{Flux_{2-79\,keV}}$\tablefootmark{\textbf{a}} & $(4.91 \pm 0.02) \times {10}^{-10}$\\
 & $\chi^2$/d.o.f. & 1184.2/1033 \\
\hline
\end{tabular}
\tablefoot{
The uncertainties are given at the 68.3\% (1$\sigma$) confidence level and were obtained using the \texttt{error} command in \textsc{xspec} with $\Delta\chi^2=1$ for one parameter of interest.
\tablefoottext{\textbf{a}}{Unabsorbed flux in units of erg\,s$^{-1}$\,cm$^{-2}$ in the 2--79\,keV range. } \tablefoottext{{p}}{The symbol $p$ denotes that the parameter hits its upper boundary. } }
\label{table:phase-avg-nustar}
\end{table}

\begin{figure}
\centering
\includegraphics[width=0.95\linewidth]{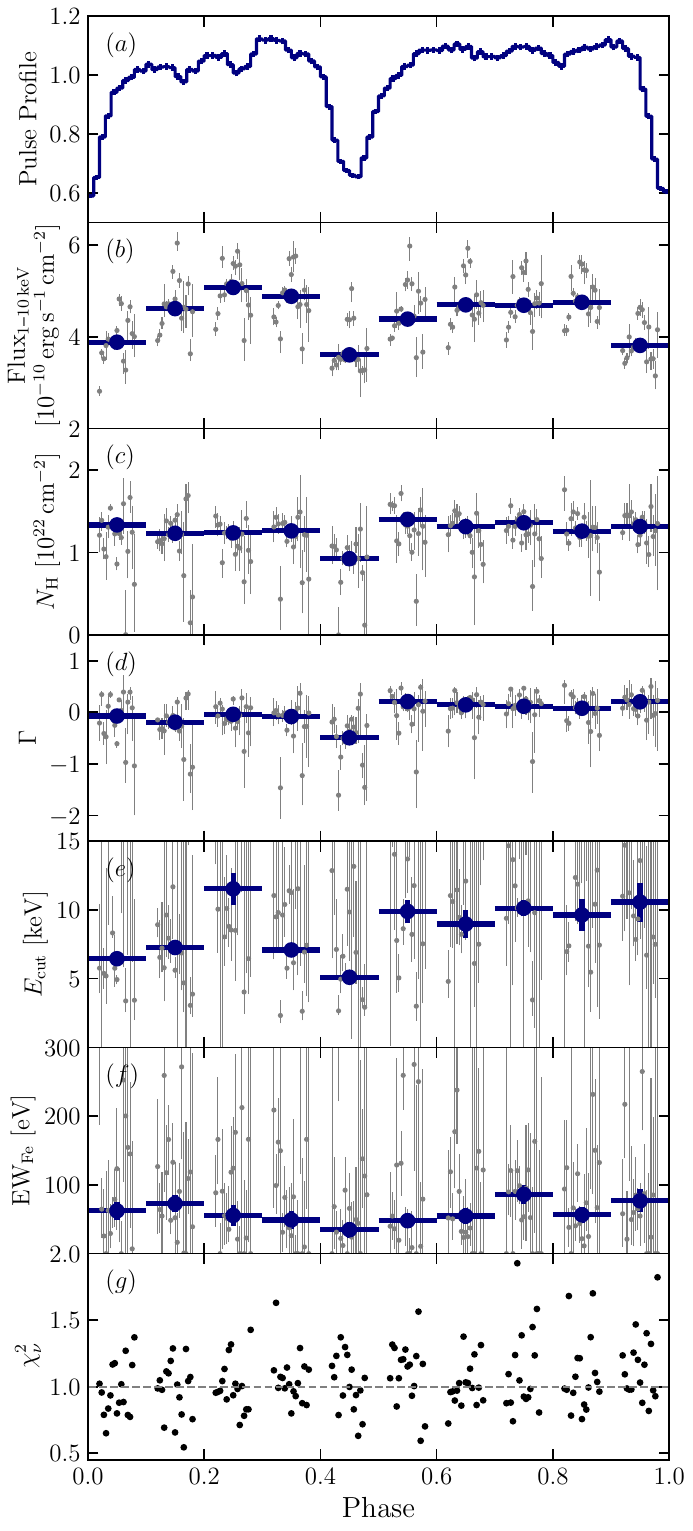}
\caption{Phase-dependent parameters of \src{} obtained from NICER phase-resolved spectral analysis. The panels show, from ($a$) to ($g$): the normalized pulse profile measured by NICER; the unabsorbed flux in 1--10\,keV; the equivalent hydrogen column density $N_\text{H}$ of the interstellar medium; the photon index $\Gamma$ and the cut-off energy $E_\mathrm{cut}$ of the continuum; the equivalent width of the iron K$\alpha$ line at 6.4\,keV; and the reduced chi-square $\chi^2_\nu$ of the best-fits of each spectrum. Grey dots indicate measurements from the phase-resolved spectral analysis of individual observations (slightly offset for clarity), while navy dots represent the average parameter values in each phase bin. }
\label{fig:phase-resolved-nicer}
\end{figure}

\subsubsection{Phase-resolved spectral analysis}

To investigate spectral variability as a function of rotational phase, we performed phase-resolved spectroscopy of the NICER and IXPE-NuSTAR data. This method enables the examination of changes in the continuum shape and line features across the pulse profile, offering insights into the geometry and emission mechanisms of the system. The pulse phase was computed using the ephemerides listed in Table~\ref{table:time-pars} according to
\begin{equation}\label{eq:phase}
\phi = \frac{1}{P} (t - t_0) - \frac{1}{2} \frac{\dot{P}}{P^2} ( t - t_0 )^2 \, ,
\end{equation}
where $P$ is the pulse period, $\dot{P}$ its time derivative at the reference epoch $t_0$, with $t_0$ also defining the zero phase. The data were divided into ten equally spaced phase bins, with source spectra and background spectra extracted using the corresponding GTIs. Each spectrum was fitted with the same model used in the phase-averaged analysis to trace the evolution of the emission components over the pulsation cycle. 

In Fig.~\ref{fig:phase-resolved-nicer}, we show the key parameters of the phase-resolved spectra, together with their average values in each phase bin, to examine their modulation with pulse phase. The decrease in the unabsorbed flux likely produces the dips observed at phases 0.0 and 0.45. The iron line is relatively weak during the NICER observation period, while the other parameters exhibit systematic modulations with pulse phase. 

For the NuSTAR observation, we primarily use the data to constrain the slope of the continuum and the strength of the narrow iron fluorescent line in the phase-resolved polarimetric analysis together with the IXPE Stokes $I$, $Q$, $U$ spectra. The results are presented in Sect.~\ref{sect:results:pol_analysis:phase-res}. 

\subsection{Polarimetric analysis}\label{sect:results:pol_analysis}

The exploratory polarimetric analysis of \src{} was carried out in \textsc{xspec}, fully accounting for both energy dispersion and the source's spectral shape. Stokes $I$, $Q$, and $U$ spectra were first extracted from all three DUs, following the methods in Sect.~\ref{sect:data:ixpe}. All nine spectra were then fitted simultaneously with \textsc{xspec}. 

The source continuum is well described by an absorbed broken power-law with a high-energy cutoff. Given the relatively soft X-ray band of IXPE, we performed a joint spectral fit with the phase-resolved NuSTAR spectra and adopted a phenomenological model that includes polarimetric information: 
\begin{align}
\texttt{constant} \times \texttt{TBabs} \times (& \texttt{bknpower} \times \texttt{highecut} \times \texttt{polconst}_{1} \notag \\
 + \ \, & \texttt{gaussian} \times \texttt{polcosnt}_{2}) \, , \notag
\end{align}
where \texttt{constant} accounts for cross-calibration between the five detectors, with the NuSTAR/FPMA factor fixed at unity. The $\texttt{polconst}_{1}$ component is used to model polarization, assuming an energy-independent PD and PA, while the PD of $\texttt{polconst}_{2}$ is fixed at 0, consistent with the assumption that the emission line is unpolarized. This setup was applied to both phase-averaged and phase-resolved analyses. 

Additionally, the model-independent \texttt{ixpepolarization} method was employed as a reference to the \textsc{xspec} results, since the two approaches are expected to produce consistent outcomes. We calculated the normalized Stokes parameters $q=Q/I$ and $u=U/I$, the PD as $\mathrm{PD} = \sqrt{q^2+u^2}$, and the PA as $\mathrm{PA} = \tfrac{1}{2}\arctan(u/q)$, measured counterclockwise from north to east on the sky \citep{Kislat_2015}. 

\subsubsection{Phase-averaged polarimetric analysis} \label{sect:results:pol_analysis:phase-avg} 

A joint phase-averaged polarimetric analysis of the \ixpe-NuSTAR observation over the full 2--79\,keV band was conducted, with the results summarized in Table~\ref{table:phase-avg-nustar}. We measured a PD of $4.8 \pm 1.2\%$ and a PA of ${17}\degr \pm {7}\degr$. Assuming a source distance of $d \simeq 7.4$\,kpc \citep{Bailer-Jones_2021}, the corresponding luminosity of \src{} at the end of the outburst is estimated as $L \simeq 3.2 \times 10^{36}$\,erg\,s$^{-1}$ in the 2--79\,keV band.

\begin{table} 
\centering
\caption{Polarimetric parameters in different energy bins for the IXPE observations with \textsc{xspec} method and \texttt{ixpepolarization} method. }
\begin{tabular}{cccccc}
    \hline\hline
    Energy & $\text{PD}_\textsc{xspec}$ & $\text{PA}_\textsc{xspec}$ & $\text{PD}_\texttt{ixpepol}$ & $\text{PA}_\texttt{ixpepol}$ \\
    $[\text{keV}]$ & $[\%]$ & $[\text{deg}]$ & $[\%]$ & $[\text{deg}]$ \\
    \hline
    2--3 & $5 \pm 3$ & $14 \pm 15$ & $6 \pm 3$ & $13 \pm 15$ \\
    3--4 & $2 \pm 2$ & $17 \pm 34$ & $3 \pm 2$ & $20 \pm 20$ \\
    4--5 & $<8$\tablefootmark{$\dagger$} & $...$ & $<6$\tablefootmark{$\dagger$} & $...$ \\
    5--6 & $14 \pm 4$ & $15 \pm 7$ & $14 \pm 4$ & $26 \pm 8$ \\
    6--8 & $8 \pm 4$ & $1 \pm 16$ & $7 \pm 4$ & $-10 \pm 20$ \\
    \hline
    2--8 & $4.8 \pm 1.2$ & $17 \pm 7$ & $4.5 \pm 1.4$ & $17 \pm 9$ \\      
    \hline
    \end{tabular}
\tablefoot{The uncertainties computed using the \texttt{error} command 
are given at the 68.3\% (1$\sigma$) confidence level ($\Delta\chi^2 = 1$ for one parameter of interest). 
\quad\tablefootmark{$(\dagger)$}{The PD is not significantly detected at the 1$\sigma$ confidence level; therefore, we report the corresponding 90\% confidence upper limit on the PD ($\Delta\chi^2 = 2.71$ for \textsc{xspec} method, and by solving the cumulative distribution function of PD for \texttt{ixpepolarization} method).}}
\label{table:pol_ebins}
\end{table}

To investigate possible spectral trends, we performed an energy-resolved polarimetric analysis by dividing the data into five intervals (2--3, 3--4, 4--5, 5--6, and 6--8\,keV). Fitting was carried out in \textsc{xspec}, using the full 2--8\,keV range for the \ixpe Stokes $I$ spectra and excluding bins outside the energy range of interest for the Stokes $Q$ and $U$ spectra. The results are presented in Table~\ref{table:pol_ebins}. For comparison, we also applied the unweighted model-independent \texttt{ixpepolarization} algorithm, which yielded broadly consistent values. 

Most energy bins show no significant polarization, with the exception of the 5--6\,keV interval, where a relatively high PD is detected. This excess is intriguing given that no additional spectral component is observed in this band, suggesting that the polarization signal may instead reflect phase-dependent variations in the PA. Although the overall phase-averaged PD is modest, the localized enhancement around 5--6\,keV points to underlying geometric or emission-related effects that warrant further investigation.

\subsubsection{Phase-resolved polarimetric analysis} \label{sect:results:pol_analysis:phase-res} 

Investigating the phase-dependent properties of pulsars is essential, as their geometry evolves with rotation. Using the ephemerides listed in Table~\ref{table:time-pars}, the phase of each photon event is calculated by Eq.~\eqref{eq:phase}. The photons were divided into ten evenly spaced phase bins, consistent with the phase-resolved analyses of NICER and NuSTAR. For each phase bin, Stokes $I$, $Q$, and $U$ spectra were generated following the procedure described in Sect.~\ref{sect:data:ixpe}. 

\begin{figure}
\centering
\includegraphics[width=0.85\linewidth]{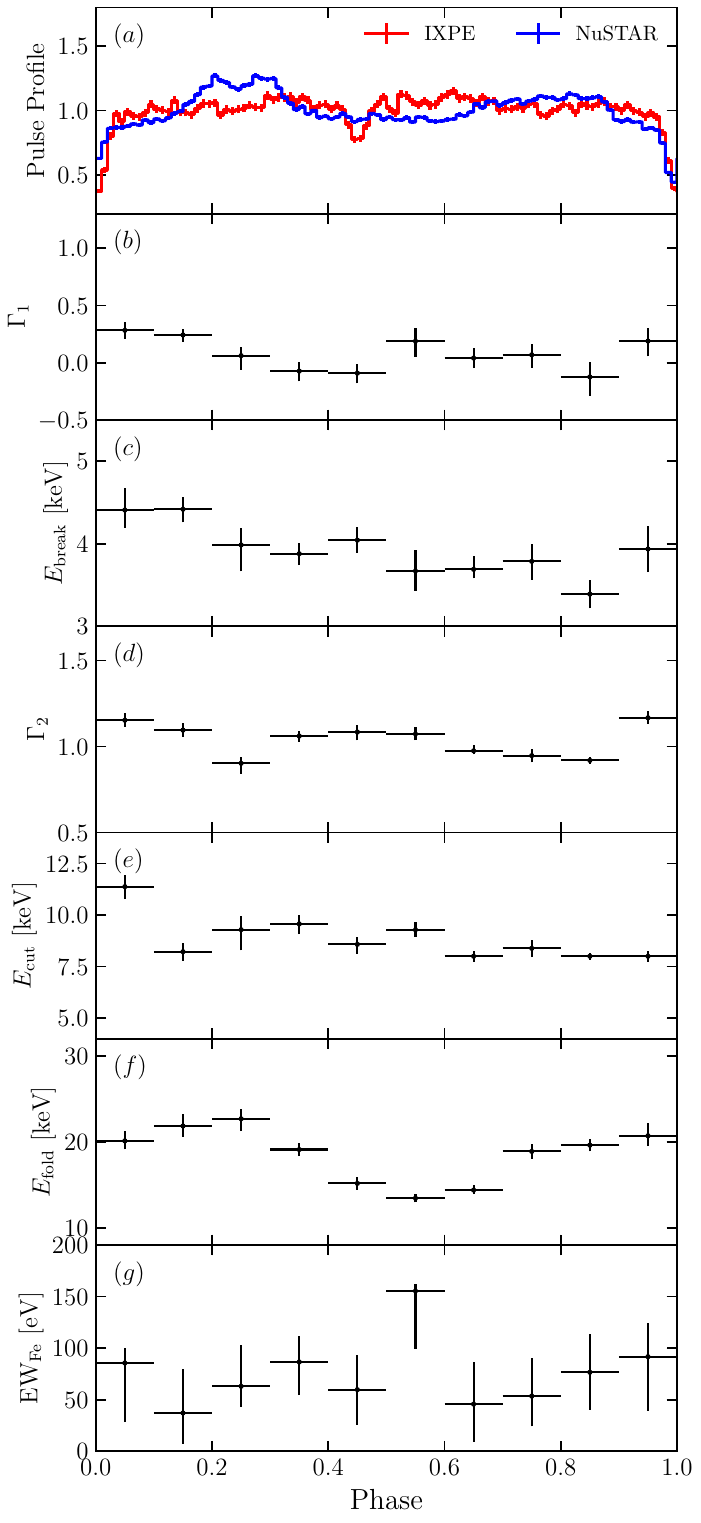}
\caption{Phase-dependent spectral parameters of \src{} derived from the joint IXPE+NuSTAR spectral-polarimetric analysis. The panels show, from ($a$) to ($g$): the normalized pulse profile, the soft photon index $\Gamma_1$, the break energy $E_\text{break}$, the hard photon index $\Gamma_2$, the cutoff energy $E_\text{cut}$ and folding energy $E_\text{fold}$ of the \texttt{highecut} component, and the equivalent width of the iron K$\alpha$ line at 6.4\,keV. }
\label{fig:phase-resolved-ixpe-nustar}
\end{figure}

\begin{figure}
\centering
\includegraphics[width=0.85\linewidth]{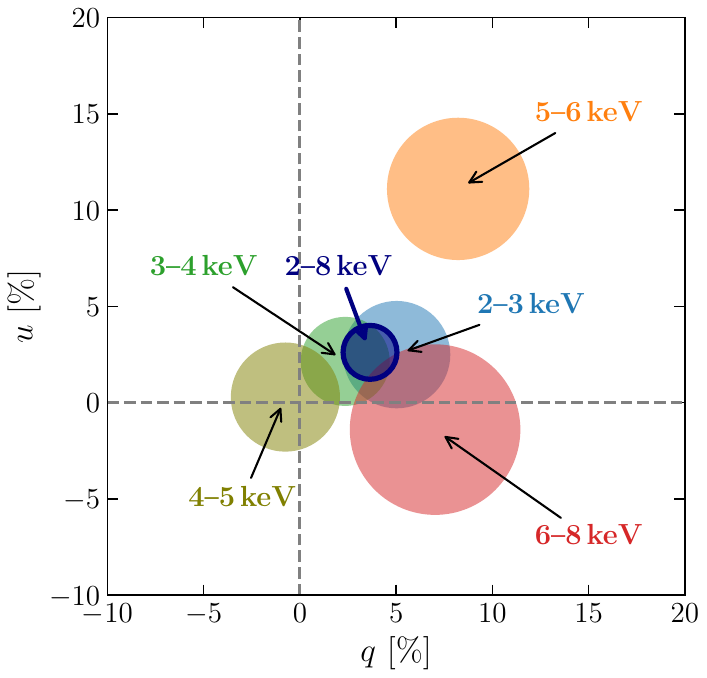}
\includegraphics[width=0.85\linewidth]{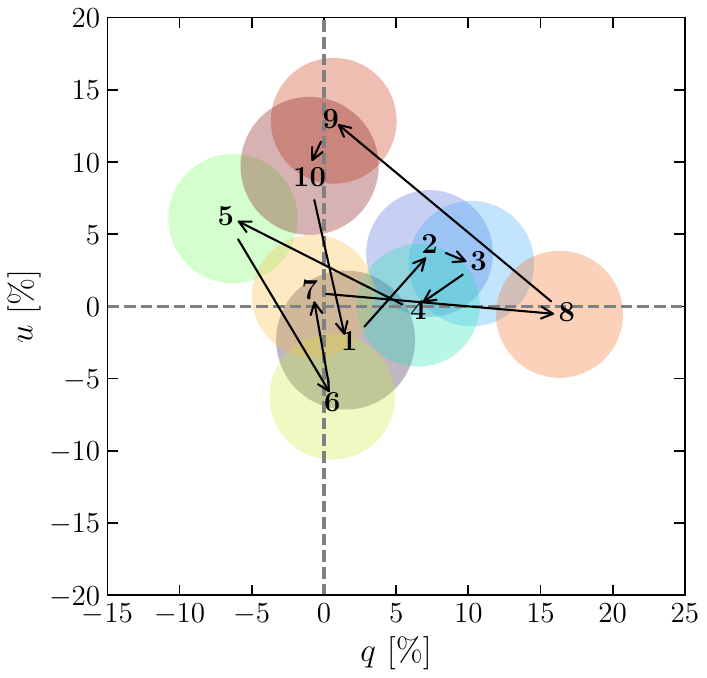}
\caption{Energy and phase dependence of the normalized Stokes parameters $q$ and $u$ during the IXPE observation, obtained using the \texttt{ixpepolarization} algorithm. For the phase-dependent analysis, the Stokes parameters were calculated in the 2--8\,keV band. The 1$\sigma$ confidence contours are shown as colored circles around each best-fit estimate. }
\label{fig:QU_src}
\end{figure}

\begin{figure}
\centering
\includegraphics[width=0.95\linewidth]{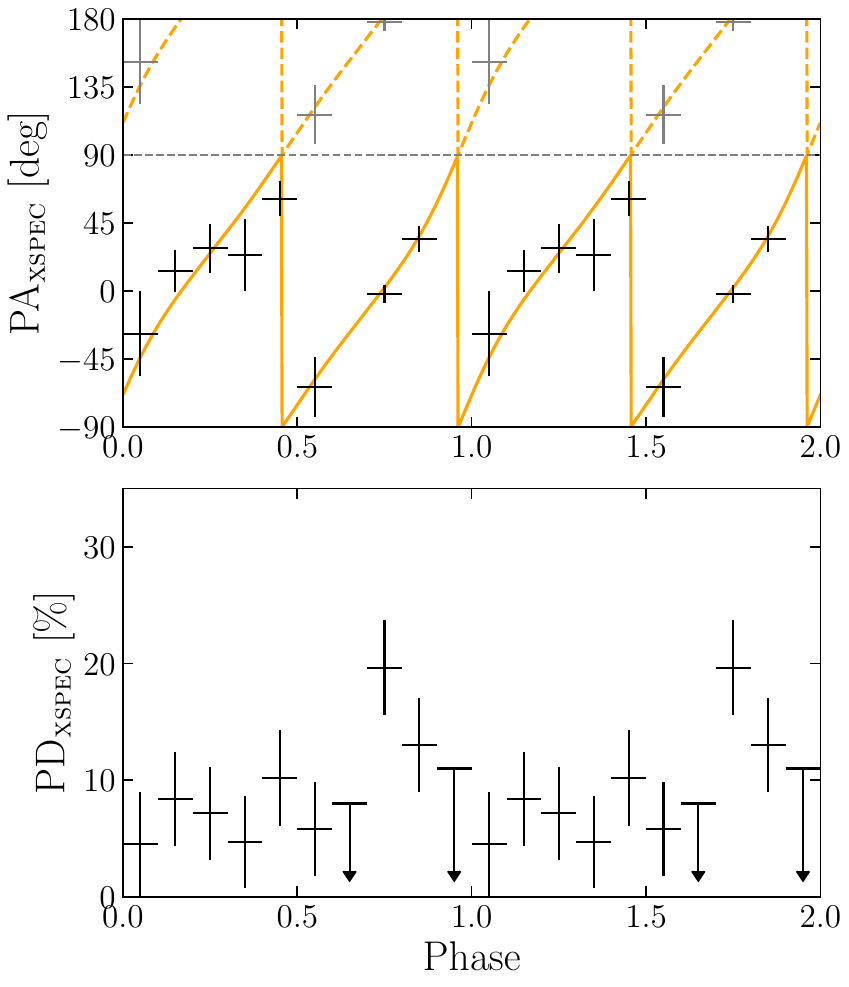}
\caption{Pulse-phase dependence of PA and PD in the 2--8\,keV band derived using the \textsc{xspec} method. For phase bins in which the PD is not significantly detected at the 1$\sigma$ confidence level, we do not report PA values; instead, we quote the 90\% upper limits on the PD. Some PAs are shown twice with an added offset of $180\degr$ to illustrate phase continuity. The orange solid curve denotes the best-fit RVM to the original PA measurements. }
\label{fig:PA_PD_RVM}
\end{figure}

We first applied the \textsc{xspec} method to each phase bin, restricting the analysis to the 2--8\,keV energy range for IXPE data and 3--79\,keV for NuSTAR. The results for the combined data set are summarized in Table~\ref{table:phase-res-xspec} and illustrated in Figs.~\ref{fig:phase-resolved-ixpe-nustar}--\ref{fig:PA_PD_RVM}. Polarization is marginally detected at a significance level exceeding $1\sigma$ in eight of the ten phase bins, providing tentative evidence for phase-dependent polarization associated with the emission geometry of the accretion column. 

\begin{figure}
\centering
\includegraphics[width=0.99\linewidth]{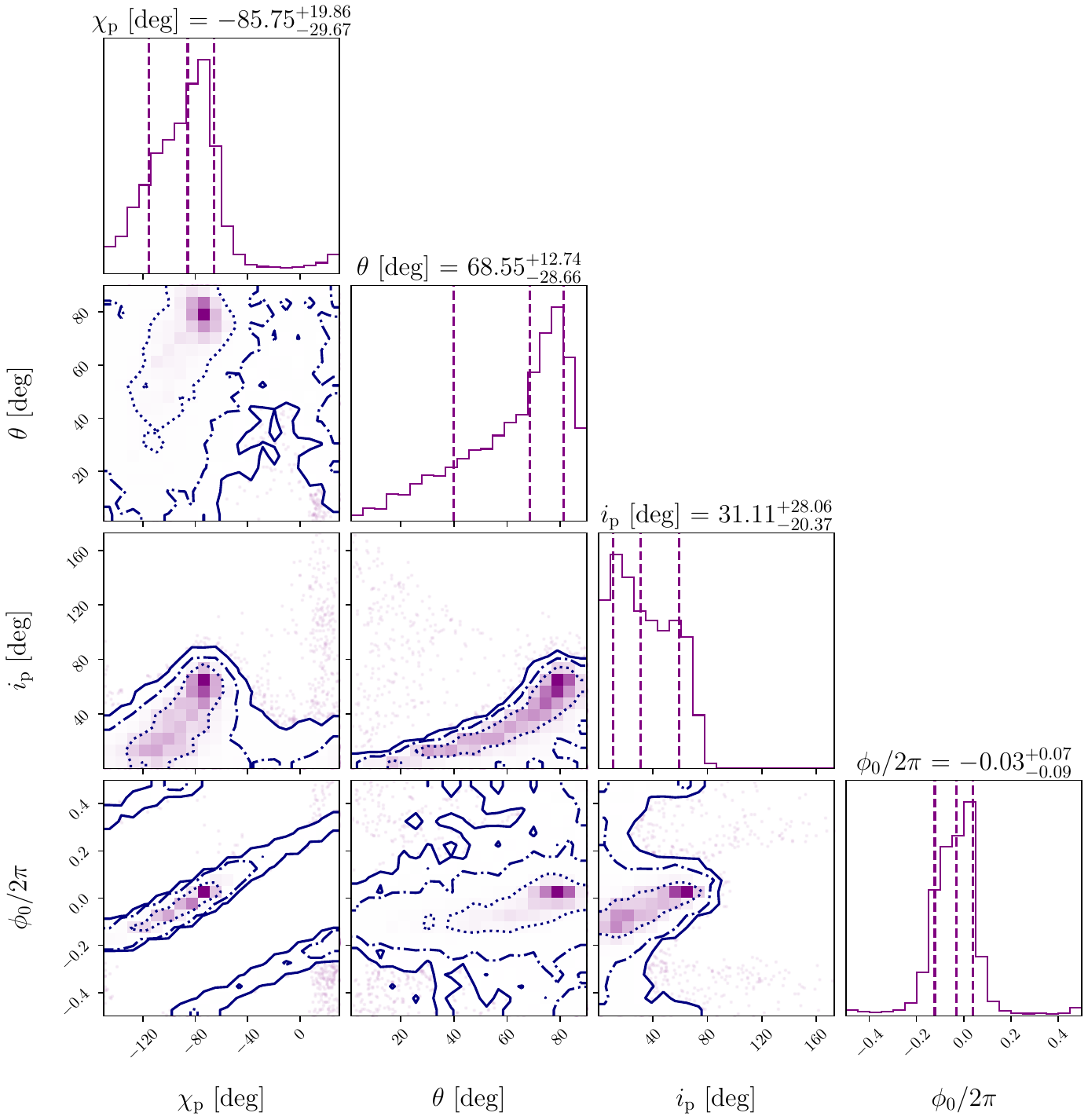}
\caption{Corner plot showing the posterior distributions of the RVM parameters describing the pulsar geometry, obtained from the PA values in the phase-resolved spectro-polarimetric analysis of the IXPE observation. The two-dimensional contours represent the 68.27\%, 95.45\%, and 99.73\% confidence levels, while the histograms display the normalized one-dimensional distributions for each parameter derived from the posterior samples. The mean values and corresponding $1\sigma$ confidence intervals are indicated above each histogram, with the dashed lines marking these intervals. }
\label{fig:corner_RVM}
\end{figure}

\begin{table*} 
\centering
\caption{Spectro–polarimetric measurement for different pulse-phase bins of \src{} from the \ixpe observations in the 2--8\,keV band, derived using \textsc{xspec} and \texttt{ixpepolarization}. }
\begin{tabular}{ccccccc}
    \hline\hline
    No. & Phase & ${\mathrm{PD}}_\textsc{xspec}$ & ${\mathrm{PA}}_\textsc{xspec}$ & $\chi^2$/d.o.f. & ${\mathrm{PD}}_\texttt{ixpepol}$ & ${\mathrm{PA}}_\texttt{ixpepol}$ \\ 
        &    & $[\%]$ & $[\text{deg}]$ &   & $[\%]$ & $[\text{deg}]$ \\
    \hline
    1 & 0.00--0.10 & $5 \pm 4$ & $-30 \pm 30$ & 641.6/681 & $<11$\tablefootmark{$\dagger$} & ... \\
    2 & 0.10--0.20 & $8 \pm 4$ & $13 \pm 14$ & 613.5/681 & $8 \pm 4$ & $13 \pm 16$ \\
    3 & 0.20--0.30 & $7 \pm 4$ & $28 \pm 17$ & 682.6/704 & $11 \pm 4$ & $8 \pm 12$ \\
    4 & 0.30--0.40 & $5 \pm 4$ & $20 \pm 30$ & 672.0/709 & $7 \pm 4$ & $0 \pm 20$ \\
    5 & 0.40--0.50 & $10 \pm 4$ & $61 \pm 12$ & 646.6/685 & $9 \pm 4$ & $68 \pm 15$ \\
    6 & 0.50--0.60 & $6 \pm 4$ & $-60 \pm 20$ & 644.9/661 & $6 \pm 4$ & $-40 \pm 20$ \\
    7 & 0.60--0.70 & $<8$\tablefootmark{$\dagger$} & ... & 663.1/696 & $<9$\tablefootmark{$\dagger$} & $...$ \\
    8 & 0.70--0.80 & $20 \pm 4$ & $-2 \pm 6$ & 673.8/698 & $16 \pm 4$ & $-1 \pm 8$ \\
    9 & 0.80--0.90 & $13 \pm 4$ & $35 \pm 9$ & 700.3/722 & $13 \pm 4$ & $43 \pm 10$ \\
    10 & 0.90--1.00 & $<11$\tablefootmark{$\dagger$} & ... & 620.4/690 & $10 \pm 5$ & $48 \pm 15$ \\
    \hline
    \end{tabular}
\tablefoot{The uncertainties in the spectro-polarimetric parameters, computed using the \texttt{error} command in \textsc{xspec}, are provided at the 68.3\% (1$\sigma$) confidence level ($\Delta\chi^2 = 1$ for one parameter of interest). 
The right two columns give PD and PA estimated with the \texttt{ixpepolarization} algorithm and the uncertainties given at the 1$\sigma$ confidence level. \ \tablefootmark{$(\dagger)$}{The PD is not significantly detected at the 1$\sigma$ confidence level; therefore, we report the corresponding 90\% confidence upper limit on the PD ($\Delta\chi^2 = 2.71$ for \textsc{xspec} method, and by solving the cumulative distribution function of PD for \texttt{ixpepolarization} method).} } 
\label{table:phase-res-xspec}
\end{table*}

The model-independent \texttt{ixpepolarization} algorithm was then applied to the phase-resolved data, and the results are presented in the last two columns of Table~\ref{table:phase-res-xspec}. In general, the polarimetric parameters derived from \texttt{ixpepolarization} are consistent with those obtained with \textsc{xspec}, even though a weighted method is adopted in \textsc{xspec} while an unweighted method is used in \texttt{ixpepolarization}. 

\section{Discussion}\label{sect:discussion}
\subsection{The rotating vector model and the geometry of the pulsar}\label{sect:discussion:RVM}
In Sect.~\ref{sect:results:pol_analysis:phase-res}, we showed that the PA varies with phase. Utilizing the polarimetric measurements in the eight detectable bins, we aim to  explore whether some geometrical parameters can be derived under the RVM \citep{Radhakrishnan_1969_RVM, Meszaros_1988_RVM, Poutanen_2020_RVM}, which has been successfully applied to several XRPs observed by \ixpe \citep[see e.g., ][]{Doroshenko_2022_HerX1, Tsygankov_2022_CenX3, Mushtukov_2023_XPersei, Doroshenko_2023_LSV, Tsygankov_2023_GROJ1008, Forsblom_2024_SMCX1}. In brief, when the emission is dominated by ordinary-mode (O-mode) photons, the RVM predicts that the PA follows Eq.~(30) from \citet{Poutanen_2020_RVM}: 
\begin{equation}\label{eq:rvm}
\tan (\text{PA} - \chi_\text{p}) = \frac{-\sin \theta \, \sin (\phi - \phi_0)}{\sin i_\text{p}\, \cos \theta - \cos i_\text{p}\,\sin \theta \, \cos (\phi - \phi_0)} \, ,
\end{equation}
where $\chi_\text{p}$ is the position angle of the pulsar angular momentum measured from north to east, $i_\text{p}$ is the inclination of the NS angular momentum to the line of sight, $\theta$ is the angle between the spin axis and the magnetic dipole, and $\phi_0$ indicates the phase when the northen magnetic pole passes in front of the observer. 

If the observed radiation is dominated by the extraordinary mode (X-mode), the position angle of the pulsar's angular momentum is given by $\chi_\text{p} \pm 90\degr$. In the non-relativistic case, the PA is independent of the polarization degree of the radiation emerging from the neutron star surface. 

The distribution of PA deviates from a normal distribution, particularly when the PD is low. Therefore, we adopted the probability density function of the PA, $\psi$, as derived by \citet{Naghizadeh-Khouei_1993}: 
\begin{equation} \label{eq:PA_dist}
G(\psi) = \frac{1}{\sqrt{\pi}}
\left\{  \frac{1}{\sqrt{\pi}}  + 
\eta {\rm e}^{\eta^2} 
\left[ 1 + {\rm erf}(\eta) \right]
\right\} {\rm e}^{-p_0^2/2} \, , 
\end{equation}
where $p_0 = \text{PD}/\text{PD}_\text{err}$ is the measured PD in units of its error, $\eta = p_0 \cos\left[ 2 \left( \psi-\psi_0 \right) \right]/\sqrt{2}$, and $\psi_0$ is the expectation value of PA. Here, \mbox{erf} denotes the error function. 

The best-fit RVM parameters are obtained by maximizing the likelihood function $\log L = \sum \log G (\psi)$. To explore the parameter space, we employ the Markov Chain Monte Carlo (MCMC) ensemble sampler implemented in the \texttt{UltraNest} package \citep{Buchner_2021_ultranest} to compute the posterior distributions of the four free RVM parameters. The resulting best-fit values from PAs derived by both the \textsc{xspec} and \texttt{ixpepolarization} methods are summarized in Table~\ref{table:RVM}. We show the corresponding covariance plots using the \textsc{xspec} method in Fig.~\ref{fig:corner_RVM} and the best-fit RVM curve to the measured PAs in Fig.~\ref{fig:PA_PD_RVM}. The parameter distributions obtained from the two methods are consistent within their respective uncertainties. 

\begin{table}
\centering
\caption{Best-fit RVM parameters for 2S~1417$-$624, derived from PAs measured with \ixpe using the \textsc{xspec}, \texttt{ixpepolarization}, and the unbinned photon-by-photon method. } 
\begin{tabular}{lccc}
\hline
\hline
Parameter & \textsc{xspec} & \texttt{ixpepol} & unbinned method\\
\hline
    $\chi_\text{p}$ [deg] & ${-90}_{-30}^{+20}$ & ${-70}_{-30}^{+70}$ & ${-100}_{-20}^{+40}$ \\
    $\theta$ [deg] & ${69}_{-29}^{+13}$ & ${77}_{-42}^{+10}$ & $50 \pm 20$ \\
    $i_\text{p}$ [deg] & ${30}_{-20}^{+30}$ & $40 \pm 40$ & ${20}_{-12}^{+17}$ \\
    $\phi_0/2\pi$ & ${-0.03}_{-0.09}^{+0.07}$ & ${0.0}_{-0.2}^{+0.3}$ & ${-0.09}_{-0.06}^{+0.10}$ \\
\hline
\end{tabular}
\label{table:RVM}
\end{table}

\begin{figure}
\centering
\includegraphics[width=0.99\linewidth]{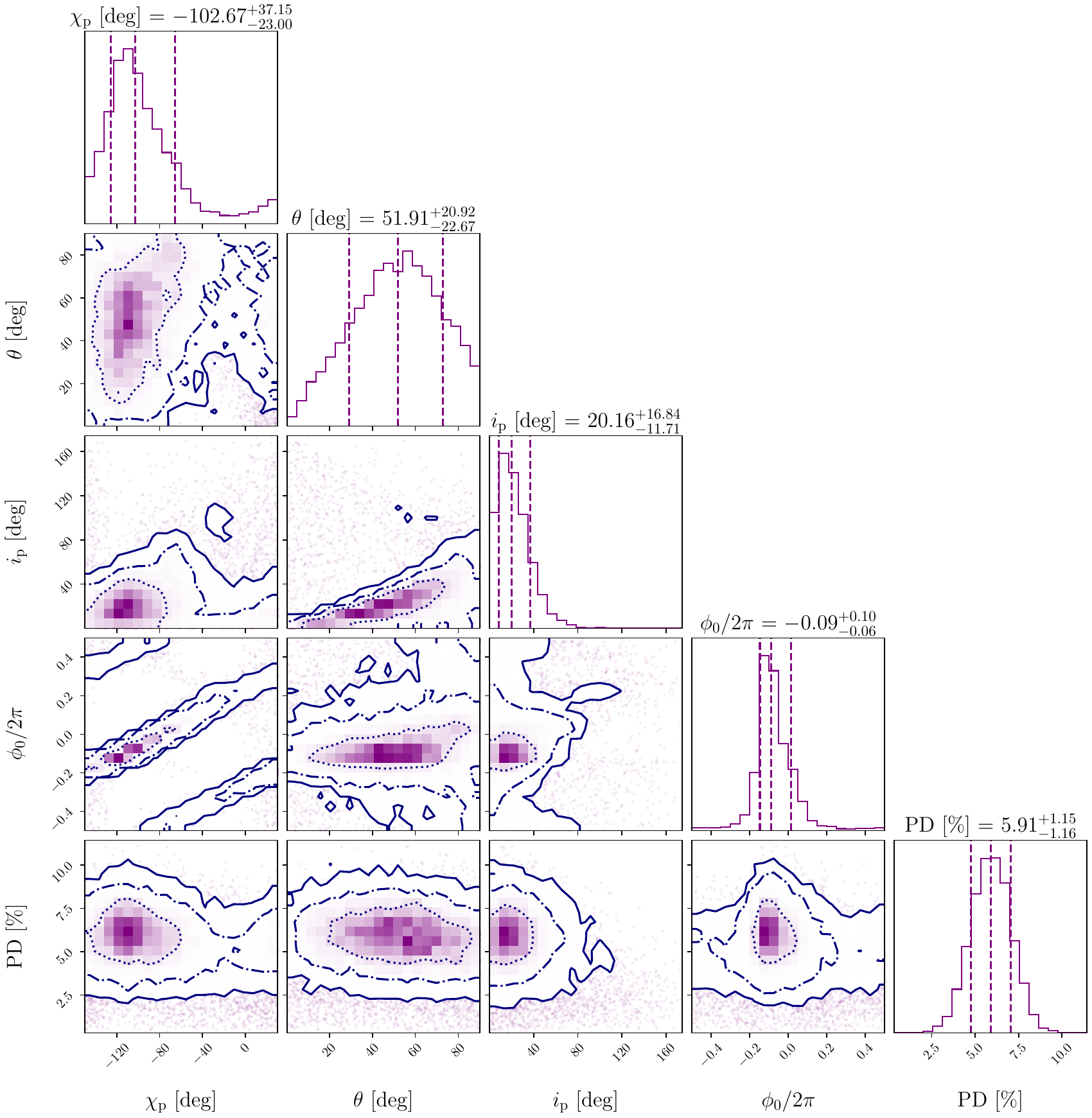}
\caption{Similar to Fig.~\ref{fig:corner_RVM}, but showing the values derived from the unbinned photon-by-photon method. }
\label{fig:corner_RVM_unbinned}
\end{figure}

We also applied the unbinned photon-by-photon method to fit the RVM \citep{Gonzalez-Caniulef_2023} (see also \citealt{Malacaria_2023_EXO2030} and \citealt{Suleimanov_2023_GX301} for examples). In this approach, we assumed a constant PD while allowing the PA to vary across the pulse phase. The results, presented in the last column of Table~\ref{table:RVM} and in Fig.~\ref{fig:corner_RVM_unbinned}, are consistent within uncertainties with those obtained from the binned PA fittings using \textsc{xspec} and \texttt{ixpepolarization}. Notably, we obtained a more significant PD of $5.9 \pm 1.2\,\%$ across the phase, corresponding to a detection significance of 5.1\,$\sigma$. This outcome is expected: variable PA naturally reduces the phase-averaged PD (with the extreme case of X~Persei; see \citealt{Mushtukov_2023_XPersei}), but when the variation of PA is accounted for, the unbinned method provides a more appropriate assessment of the overall detection significance than the phase-averaged PD reported in Sect.~\ref{sect:results:pol_analysis:phase-avg}. We caution, however, that the instrument's effective area is not considered in this approach. Since the effective area of IXPE peaks at $\sim$2\,keV, most of the detected photons are concentrated in this energy range, which may introduce bias into the fit. 

Although subject to considerable uncertainties, the RVM fit suggests that \src{} possesses a large magnetic obliquity, compatible with a quasi-orthogonal geometry, though smaller obliquities cannot be entirely ruled out. The $\phi_0$ angle, which denotes the phase at which the northern magnetic pole comes closest to the observer, is well constrained to be near phase zero, coinciding with the main dip in the pulse profile. Such a geometry approaches that of an orthogonal rotator, an extreme case of pulsar configuration that gives rise to the most pronounced emission and polarization signatures, and is therefore of fundamental importance for probing the magnetic field structure, radiation mechanisms, and geometrical evolution of neutron stars. 

\subsection{A more detailed polarimetric investigation in the 5--6\,keV band}

As shown in Sect.~\ref{sect:results:pol_analysis:phase-avg}, the 5--6\,keV band exhibits a notably stronger polarization than the other energy bands. This may indicate either that photons in this band have an intrinsically higher PD while following a PA modulation similar to that at other energies, or that the PA modulation itself becomes weaker or changes with energy in the 5--6\,keV band. To investigate this possibility, we follow the approach of \citet{Zhao_2026} and divide the IXPE energy range into three bands: 2--5\,keV, 5--6\,keV, and 6--8\,keV. 

\begin{figure*}
\centering
\includegraphics[width=0.95\linewidth]{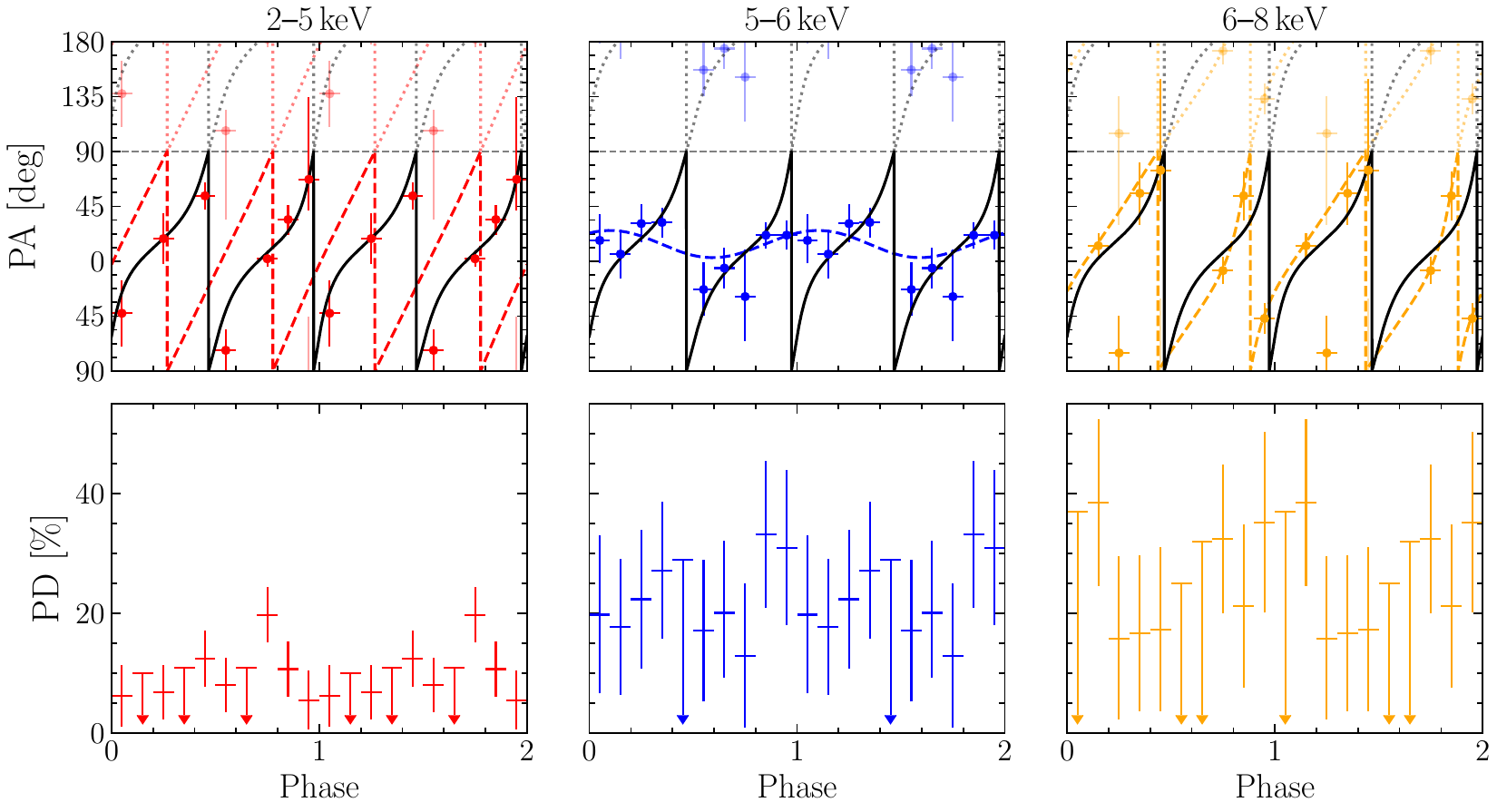}
\caption{Pulse-phase-resolved PAs and PDs in different energy bands obtained using the \textsc{xspec} method. For phase bins in which the PD is not significantly detected at the 1$\sigma$ confidence level, we do not report PA values; instead, we quote the 90\% upper limits on the PD. Some PA values have been shifted by 180$\degr$ to better illustrate their phase continuity. The colored dashed curve in each panel represents the best-fit RVM to the PA measurements in the corresponding energy band, while the black solid curves show the simultaneous best-fit RVM to the Stokes parameters across the three energy bands. }
\label{fig:PD_PA_phase-res-xspec-energy}
\end{figure*}

We then used the \textsc{xspec} method to measure the phase-resolved PAs and PDs in each energy band. The results are shown in Table~\ref{table:phase-res-xspec-energy} and Fig.~\ref{fig:PD_PA_phase-res-xspec-energy}. We find some PA discrepancies between different energy bands in individual phase bins. The largest difference occurs in the phase interval 0.90--1.00, where the PAs in the 5--6\,keV and 6--8\,keV bands differ by $\Delta\mathrm{PA} \simeq 68^\circ$, corresponding to a significance of $3.8\,\sigma$. The second largest discrepancy is found in the phase interval 0.20--0.30, between the 2--5\,keV and 6--8\,keV bands, with $\Delta\mathrm{PA}\simeq90^\circ$, corresponding to $2.2\,\sigma$. 

\begin{table*}
\centering
\caption{Phase-resolved spectro-polarimetric measurements of \src{} in different energy bands, derived using \textsc{xspec} method.}
\begin{tabular}{cccccccc}
    \hline\hline
    No. & Phase 
    & \multicolumn{2}{c}{2--5\,keV}
    & \multicolumn{2}{c}{5--6\,keV}
    & \multicolumn{2}{c}{6--8\,keV} \\
        & 
    & PD [\%] & PA [deg]
    & PD [\%] & PA [deg]
    & PD [\%] & PA [deg] \\
    \hline
    1  & 0.00--0.10 
    & $6 \pm 5$ & $-40 \pm 30$
    & $20 \pm 13$ & $20 \pm 20$
    & $<37$\tablefootmark{$\dagger$} & ... \\

    2  & 0.10--0.20 
    & $<10$\tablefootmark{$\dagger$} & ...
    & $18 \pm 11$ & $10 \pm 20$
    & $39 \pm 14$ & $13 \pm 11$ \\

    3  & 0.20--0.30 
    & $7 \pm 5$ & $20 \pm 20$
    & $22 \pm 12$ & $31 \pm 16$
    & $16 \pm 14$ & $-80 \pm 30$ \\

    4  & 0.30--0.40 
    & $<11$\tablefootmark{$\dagger$} & ...
    & $27 \pm 11$ & $32 \pm 12$
    & $17 \pm 13$ & $60 \pm 30$ \\

    5  & 0.40--0.50 
    & $12 \pm 5$ & $54 \pm 11$
    & $<29$\tablefootmark{$\dagger$} & ...
    & $17 \pm 14$ & $70 \pm 30$ \\

    6  & 0.50--0.60 
    & $8 \pm 5$ & $-73 \pm 17$
    & $17 \pm 12$ & $-20 \pm 20$
    & $<25$\tablefootmark{$\dagger$} & ... \\

    7  & 0.60--0.70 
    & $<11$\tablefootmark{$\dagger$} & ...
    & $20^{+12}_{-11}$ & $-6 \pm 17$
    & $<32$\tablefootmark{$\dagger$} & ... \\

    8  & 0.70--0.80 
    & $20 \pm 5$ & $2 \pm 7$
    & $13 \pm 12$ & $-30 \pm 40$
    & $32 \pm 12$ & $-8 \pm 11$ \\

    9  & 0.80--0.90 
    & $11 \pm 5$ & $34 \pm 12$
    & $33 \pm 12$ & $21 \pm 11$
    & $21 \pm 14$ & $50 \pm 20$ \\

    10 & 0.90--1.00 
    & $6 \pm 5$ & $70 \pm 30$
    & $31 \pm 13$ & $21 \pm 12$
    & $35 \pm 15$ & $-47 \pm 13$ \\
    \hline
\end{tabular}
\tablefoot{The uncertainties in the spectro-polarimetric parameters, computed using the \texttt{error} command in \textsc{xspec}, are provided at the 68.3\% (1$\sigma$) confidence level ($\Delta\chi^2 = 1$ for one parameter of interest). 
\tablefootmark{$(\dagger)$}{The PD is not significantly detected at the 1$\sigma$ confidence level; therefore, we report the corresponding 90\% confidence upper limit on the PD ($\Delta\chi^2 = 2.71$ for \textsc{xspec} method). }}
\label{table:phase-res-xspec-energy}
\end{table*}

\begin{figure}
\centering
\includegraphics[width=0.99\linewidth]{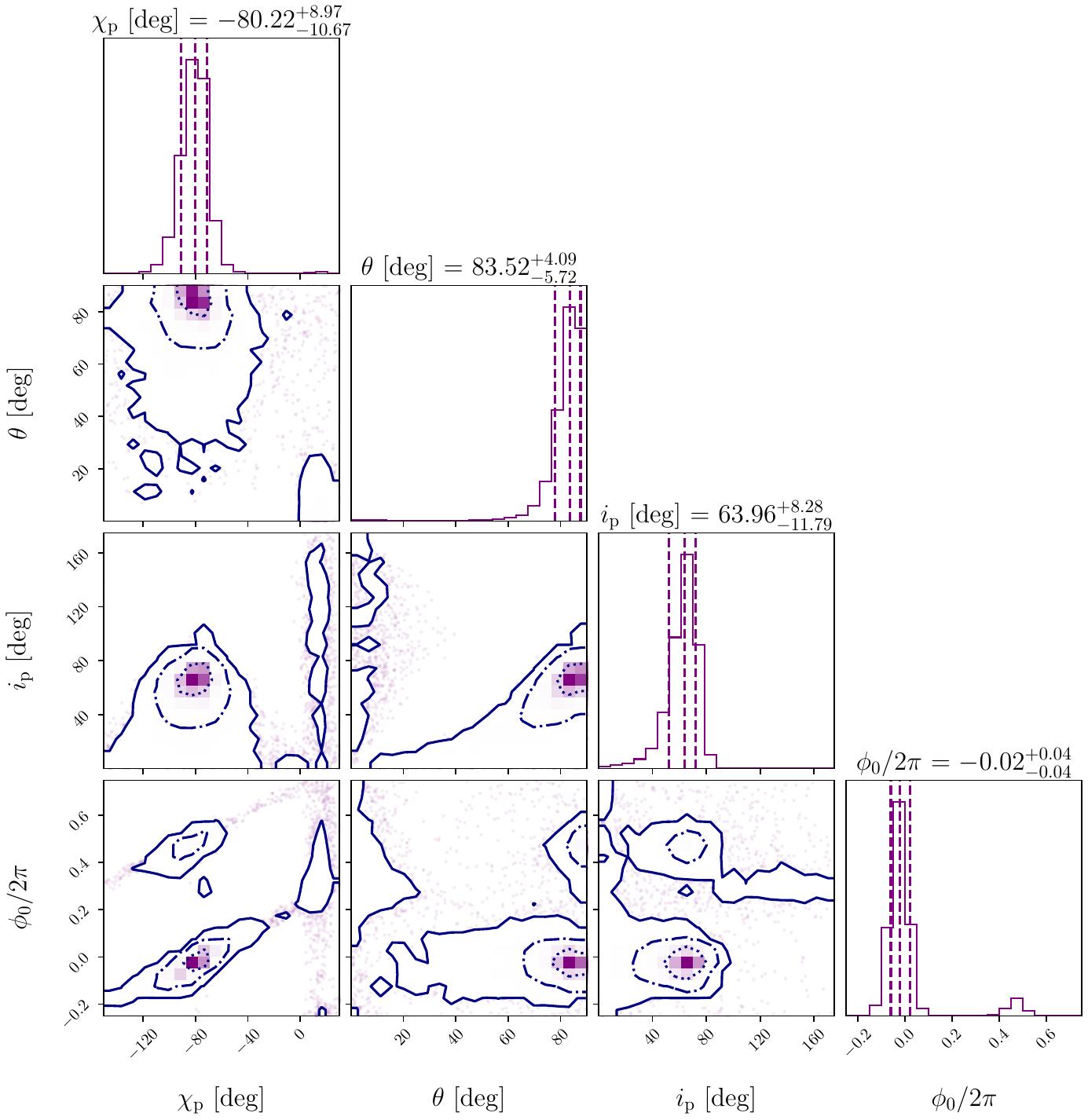}
\caption{Similar to Fig.~\ref{fig:corner_RVM}, but showing the values derived from the cross-fitting of the Stokes Q/U parameters in the three energy bands (2--5\,keV, 5--6\,keV, and 6--8\,keV). }
\label{fig:corner_RVM_QU}
\end{figure}

We fitted the RVM independently in the three energy bands, but found that the resulting parameter constraints are rather weak (see Table~\ref{table:RVM-energy}). We also attempted a simultaneous fit to the Stokes parameters Q/U across the three energy bands (see Table~\ref{table:phase-res-qu-energy} in the Appendix), without including an additional constant polarized component, and obtained $\chi^2/\mathrm{d.o.f.}=26.8/56$. This fit yields the strongest constraints on the RVM parameters, as shown in the last column of Table~\ref{table:RVM-energy} and in Fig.~\ref{fig:corner_RVM_QU}. The magnetic obliquity $\theta$ is tightly constrained to be close to a right angle, while the inclination of the neutron-star spin axis is approximately $60\degr$. However, the very low reduced $\chi^2$ suggests that the fit is likely overfitting the data. For this reason, we did not include an additional constant polarized component in our model. In addition, Fig.~\ref{fig:PD_PA_phase-res-xspec-energy} shows that the PAs measured in the individual energy bands remain compatible with a common RVM model across all three bands, although the resulting PA modulation may differ from the best-fit RVM curves obtained for the individual bands. 

\begin{table*}
\centering
\caption{Best-fit RVM parameters for \src{} in different energy bands, derived from the PAs measured with \ixpe using the \textsc{xspec} method. } 
\begin{tabular}{lcccc}
\hline
\hline
Parameter & 2--5\,keV & 5--6\,keV & 6--8\,keV & Cross-fitting \\
\hline
    $\chi_\text{p}$ [deg] & ${-60}_{-40}^{+60}$ & ${13}_{-9}^{+7}$ & ${-113}_{-15}^{+19}$ & ${-80}_{-11}^{+9}$ \\
    $\theta$ [deg] & ${73}_{-29}^{+15}$ & ${13}_{-8}^{+15}$ & ${43} \pm 19$ & ${84}_{-6}^{+4}$ \\
    $i_\text{p}$ [deg] & ${14}_{-11}^{+23}$ & ${80}_{-30}^{+40}$ & ${28}_{-15}^{+21}$ & ${64}_{-12}^{+8}$ \\
    $\phi_0/2\pi$ & $-0.2 \pm 0.3$ & ${0.39} \pm 0.13$ & ${-0.14} \pm 0.04$ & $-0.02 \pm 0.04$ \\
\hline
\end{tabular}
\tablefoot{
The last column also reports the simultaneous best-fit RVM to the Stokes parameters Q/U across the three energy bands. 
}
\label{table:RVM-energy}
\end{table*}

The PD values in the 5--6\,keV band are higher than those in the 2--5\,keV band in most phase bins. However, due to the large statistical uncertainties, these differences are not significant in individual phase bins. The enhanced polarization in the 5--6\,keV band may be related to this interval sampling a relatively clean continuum window, whereas the broader bands may be more strongly diluted by additional weakly polarized or unpolarized components, such as reprocessed emission and fluorescent lines. At the same time, the PA differences seen in some phase bins suggest that simple dilution alone may not fully explain the broadband polarization behavior. Instead, the data may require the superposition of multiple components with distinct polarization properties. 

The observed energy dependence is also likely coupled to the pulse phase. This implies that the relative contributions of the different emission and reprocessing components change as the neutron star rotates, naturally producing phase-dependent variations in both PD and PA. In this picture, the polarization measurements indicate that the observed radiation in the 2--8\,keV band does not originate from a single homogeneous emission region, but instead reflects a combination of direct accretion-column emission and additional phase-dependent reprocessed components.

At present, few attempts have been made to explain the relatively smooth energy dependence of the PA in XRPs at photon energies well below the cyclotron resonance energy \citep[e.g.,][]{Forsblom_2025_VelaX1}. Consequently, an abrupt PA change between two nearby energy bands in the soft X-ray regime is not straightforward to interpret within the current theoretical framework. A comparable phenomenon has been reported in 4U~1907+09 \citep{Zhou_2025_4U1907}, where a PA swing of approximately $90\degr$ was marginally detected between two adjacent energy bands within a single phase bin. It is worth noting that both cases were observed under photon-starved conditions, where limited photon statistics reduce the robustness of the measurements. Recent theoretical and observational studies have also emphasized that complex mixtures of emission components and reprocessing regions may introduce additional energy-dependent polarization signatures in accreting XRPs \citep[e.g.,][]{Zhao_2026}. Future observations with significantly improved photon statistics and broader energy coverage will therefore be crucial for confirming these features and for providing the empirical basis needed to develop a physical explanation. 

\subsection{Possible scenario for the 2025 outburst of 2S~1417$-$624}

The observed flux from NuSTAR is $\sim 4.91 \times 10^{-10}$\,erg\,s$^{-1}$\,cm$^{-2}$ (in the 2--79\,keV band), which corresponds to a luminosity of $3.2 \times 10^{36}$\,erg\,s$^{-1}$, assuming a source distance of $7.4\,\mathrm{kpc}$ \citep{Bailer-Jones_2021}. 
\citet{LiuQ_2024} reported a possible cyclotron line at $100\,\mathrm{keV}$ in this source, implying a magnetic field strength of $\sim 9 \times {10}^{12}$\,G. 
Given that accreting pulsars can operate in different radiative regimes depending on their luminosity, the critical luminosities of this source are estimated to be $\sim1.55 \times 10^{37}$\,erg\,s$^{-1}$ and $\sim 5 \times 10^{37}$\,erg\,s$^{-1}$, as determined using the approaches of \citet{Becker_2012} and \citet{Mushtukov_2015}, respectively. 
Since the NuSTAR observation occurred at the end of the outburst, the corresponding MAXI count rate was $\sim$ $0.025$ cts/s, whereas the outburst peak reached a MAXI count rate of $\sim 0.1$ cts/s, four times higher than during the NuSTAR observation. 
Therefore, the peak luminosity of the 2025 outburst can be roughly estimated as four times the NuSTAR luminosity:
$L_{\mathrm{peak}} \approx 4 \times L_{\mathrm{NuSTAR}}$ = $1.3 \times 10^{37}$\,erg\,s$^{-1}$. It is shown that even the peak luminosity did not exceed the critical luminosity; therefore, the radiation regime remained in the pencil-beam pattern throughout the entire outburst. The absence of significant luminosity-dependent changes in the pulse profile (as observed with NICER, IXPE, and NuSTAR) further supports this conclusion.  
We also note that the distance to this source may carry some uncertainty, which in turn affects the luminosity estimate. \cite{LiuQ_2025} reported two characteristic flux levels associated with changes in the pulse profile, at $\sim 4.1 \times 10^{-9}$ and $6.4 \times 10^{-9}\,\mathrm{erg\,cm^{-2}\,s^{-1}}$ (2--140 keV). Below $4.1 \times 10^{-9}\,\mathrm{erg\,cm^{-2}\,s^{-1}}$, the emission is thought to be dominated by a pencil-beam pattern. Therefore, even though the peak flux of the 2025 outburst is about $F_{\mathrm{peak}} \approx 2.0 \times 10^{-9}\,\mathrm{erg\,cm^{-2}\,s^{-1}}$ (3--79 keV, equivalent to $4 \times F_{\mathrm{NuSTAR}}$), it still lies below the threshold at which a transition in the radiation pattern is expected. (The difference in flux integration energy bands does not significantly affect this conclusion, as the broader energy band is not expected to further increase the measured flux substantially.)
Polarization measurements, which constrain the geometry of accreting pulsars, greatly enhance our understanding of the pulse profile. 
With $\theta \approx 70\degr$, as determined by fitting PAs obtained by \textsc{xspec} and \texttt{ixpepolarization}, and also the unbinned method, the source is probably close to a quasi-orthogonal rotator configuration. Considering $i_\text{p} \approx 24\degr$ from the same methods, and the relatively wide beam angle of a pencil-beam emission pattern \citep{Hu_2023}, it is plausible to observe two flux maxima within a single rotational cycle, consistent with the observed pulse profile. 

\section{Conclusions}\label{sect:summary}

The multi-mission campaign on 2S~1417$-$624 during its weak type-II outburst in 2025 has provided new insights into the source's spectral and polarimetric properties. Phase-averaged and phase-resolved spectroscopy with NICER and NuSTAR shows that the spectra are well described by a broken power-law with a high-energy cut-off, with several parameters exhibiting systematic pulse-phase modulations indicative of changes in emission geometry. With IXPE, we obtained the first polarimetric measurements of this source, yielding a phase-averaged polarization degree of $4.8 \pm 1.2\%$ and a polarization angle of ${17} \pm {7}\degr$. By applying the unbinned photon-by-photon method and accounting for PA variation across the pulse phase, we obtained a PD of $5.9\pm 1.2 \,\%$, corresponding to a detection significance of 5.1\,$\sigma$, thereby reinforcing the robustness of our polarimetric results. Complementary phase-resolved polarimetric analysis suggests that 2S~1417$-$624 possesses a magnetic obliquity of $\theta = 69_{-29}^{+13}$\,deg, indicating that the geometry is compatible with a quasi-orthogonal rotator within uncertainties. The tightest constraints on the geometrical parameters can be obtained from a simultaneous RVM fit to the Stokes Q/U parameters of the three energy bands, 2--5\,keV, 5--6\,keV, and 6--8\,keV, which gives $\theta = {84}_{-6}^{+4}$\,deg. Together, these results establish a consistent picture in which both the spectral and polarization properties vary significantly over the pulse cycle, offering important constraints on the geometry and emission processes in the accretion region of this transient XRP. 

\begin{acknowledgements}
This work reports observations obtained with the Imaging X-ray Polarimetry Explorer (IXPE), a joint US (NASA) and Italian (ASI) mission, led by Marshall Space Flight Center (MSFC). The research uses data products provided by the IXPE Science Operations Center (MSFC), using algorithms developed by the IXPE Collaboration, and distributed by the High-Energy Astrophysics Science Archive Research Center (HEASARC). 
This research has made use of data from the NuSTAR mission, a project led by the California Institute of Technology, managed by the Jet Propulsion Laboratory, and funded by the National Aeronautics and Space Administration. Data analysis was performed using the NuSTAR Data Analysis Software (NuSTARDAS), jointly developed by the ASI Science Data Center (SSDC, Italy) and the California Institute of Technology (USA).
This research has been supported by the Deutsche Forschungsgemeinschaft (DFG) grants 549824807 (LD) and WE 1312/59-1 (VFS), and the UKRI Stephen Hawking fellowship (AAM). 
P.-J.W. is grateful for the financial support provided by the Sino-German (CSC-DAAD) Postdoc Scholarship Program (57678375). This research was supported by the International Space Science Institute (ISSI) in Bern, through International Team project 25-657 'Polarimetric Insights into Extreme Magnetism’. SST and JP acknowledge support by the Research Council of Finland Centre of Excellence in Neutron-Star Physics (project 374064). Q. L. acknowledges support by the Science Foundation of Hebei Normal University (No.~L2026B49), Top-notch Talent Project (Overseas Platform) under the Hebei Yanzhao Golden Terrace Talent Gathering Program. 
\end{acknowledgements}

\bibliographystyle{yahapj}
\bibliography{references}

\onecolumn
\begin{appendix}
\section{The phase-resolved Stokes parameters of the 2--5\,keV, 5--6\,keV, and 6--8\,keV band}

\begin{table*}[h!]
\centering
\caption{Phase-resolved Stokes parameters of \src{} in different energy bands, derived with the \texttt{ixpepolarization} method using the \texttt{NEFF} weighting scheme.}
\begin{tabular}{cccccccc}
    \hline\hline
    No. & Phase 
    & \multicolumn{2}{c}{2--5\,keV}
    & \multicolumn{2}{c}{5--6\,keV}
    & \multicolumn{2}{c}{6--8\,keV} \\
        & 
    & $Q$ & $U$
    & $Q$ & $U$
    & $Q$ & $U$ \\
    \hline
    1  & 0.00--0.10 
    & $10 \pm 70$   & $-60 \pm 70$
    & $20 \pm 30$   & $10 \pm 30$
    & $-10 \pm 20$  & $10 \pm 20$ \\

    2  & 0.10--0.20 
    & $50 \pm 80$   & $30 \pm 80$
    & $40 \pm 30$   & $10 \pm 30$
    & $60 \pm 30$   & $20 \pm 30$ \\

    3  & 0.20--0.30 
    & $60 \pm 80$   & $60 \pm 80$
    & $40 \pm 30$   & $30 \pm 30$
    & $-20 \pm 30$  & $-40 \pm 30$ \\

    4  & 0.30--0.40 
    & $30 \pm 80$   & $-10 \pm 80$
    & $30 \pm 30$   & $80 \pm 30$
    & $0 \pm 30$    & $10 \pm 30$ \\

    5  & 0.40--0.50 
    & $-50 \pm 80$  & $90 \pm 80$
    & $-20 \pm 30$  & $-20 \pm 30$
    & $-10 \pm 20$  & $0 \pm 20$ \\

    6  & 0.50--0.60 
    & $-40 \pm 80$  & $-50 \pm 80$
    & $30 \pm 30$   & $-30 \pm 30$
    & $10 \pm 20$   & $0 \pm 20$ \\

    7  & 0.60--0.70 
    & $-10 \pm 80$  & $20 \pm 80$
    & $50 \pm 30$   & $0 \pm 30$
    & $20 \pm 30$   & $-10 \pm 30$ \\

    8  & 0.70--0.80 
    & $210 \pm 80$  & $0 \pm 80$
    & $0 \pm 30$    & $-10 \pm 30$
    & $70 \pm 20$   & $10 \pm 20$ \\

    9  & 0.80--0.90 
    & $30 \pm 80$   & $140 \pm 80$
    & $30 \pm 30$   & $70 \pm 30$
    & $0 \pm 20$    & $30 \pm 20$ \\

    10 & 0.90--1.00 
    & $-40 \pm 70$  & $60 \pm 70$
    & $20 \pm 30$   & $30 \pm 30$
    & $0 \pm 20$    & $-30 \pm 20$ \\
    \hline
\end{tabular}
\tablefoot{The uncertainties of the Stokes parameters are provided at the 68.3\% confidence level. This approach can lead to differences between the PAs and PDs derived from the Stokes parameters and those reported in Table~\ref{table:phase-res-xspec-energy}. }
\label{table:phase-res-qu-energy}
\end{table*}
\end{appendix}
\end{document}